\documentclass[journal=ancac3,manuscript=article]{achemso}

\usepackage{graphicx} 
\usepackage{amsmath} 
\usepackage[version=4]{mhchem}
\usepackage{color}
\usepackage{footnote}

\usepackage{xr}
\makeatletter

\newcommand*{\addFileDependency}[1]{
\typeout{(#1)}
\@addtofilelist{#1}
\IfFileExists{#1}{}{\typeout{No file #1.}}
}\makeatother

\newcommand*{\myexternaldocument}[1]{%
\externaldocument{#1}%
\addFileDependency{#1.tex}%
\addFileDependency{#1.aux}%
}

\myexternaldocument{supporting}

\usepackage[hidelinks]{hyperref}
\hypersetup{
  colorlinks   = true, 
  urlcolor     = blue, 
  linkcolor    = red, 
  citecolor   = blue 
}
\usepackage{siunitx}
\usepackage{lineno}

\newcommand{\LVO}{LaVO$_3$}
\newcommand{\DSO}{DyScO$_3$}
\newcommand{\GSO}{GdScO$_3$}

\newcommand{\CTO}{CaTiO$_3$}
\newcommand{\LMO}{LaMnO$_3$}

\DeclareSIUnit\angstrom{\text {Å}}

\author{Duncan T.L. Alexander}
\email{duncan.alexander@epfl.ch}
\affiliation{Electron Spectrometry and Microscopy Laboratory (LSME), Institute of Physics (IPHYS), École Polytechnique Fédérale de Lausanne (EPFL), Lausanne, Switzerland}
\author{Hugo Meley}
\affiliation{Department of Quantum Matter Physics (DQMP), University of Geneva, Geneva, Switzerland}
\author{Michael Marcus Schmitt}
\affiliation{Physique Théorique des Matériaux, Université de Liège (B5), Liège, Belgium}
\author{Bernat Mundet}
\affiliation{Department of Quantum Matter Physics (DQMP), University of Geneva, Geneva, Switzerland}
\alsoaffiliation{Electron Spectrometry and Microscopy Laboratory (LSME), Institute of Physics (IPHYS), École Polytechnique Fédérale de Lausanne (EPFL), Lausanne, Switzerland}
\author{Philippe Ghosez}
\affiliation{Physique Théorique des Matériaux, Université de Liège (B5), Liège, Belgium}
\author{Jean-Marc Triscone}
\affiliation{Department of Quantum Matter Physics (DQMP), University of Geneva, Geneva, Switzerland}
\author{Stefano Gariglio}
\email{stefano.gariglio@unige.ch}
\affiliation{Department of Quantum Matter Physics (DQMP), University of Geneva, Geneva, Switzerland}

\title{Engineering Symmetry Breaking Interfaces by Nanoscale Structural--Energetics in Orthorhombic Perovskite Thin Films}


\begin{document}

\begin{tocentry}

\includegraphics[width=8.3cm]{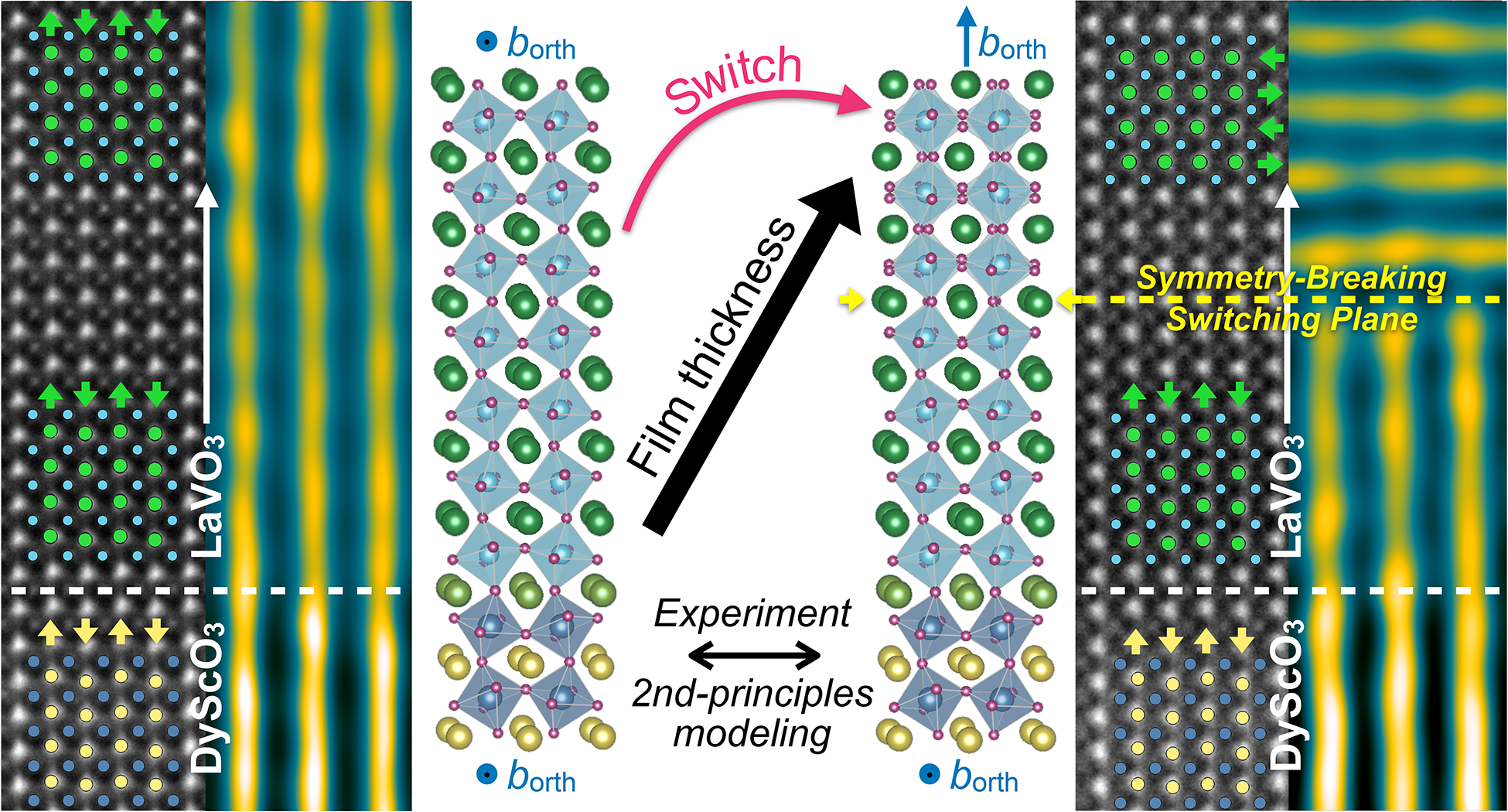}

\end{tocentry}

\begin{abstract}

The atomic configuration of phases and their interfaces is fundamental to materials design and engineering. Here, we unveil a transition metal oxide interface, whose formation is driven by energetic influences---epitaxial tensile strain versus oxygen octahedra connectivity---that compete in determining the orientation of an orthorhombic perovskite film. We study this phenomenon in a system of \LVO{} grown on $(101)$ \DSO{}, using atomic-resolution scanning transmission electron microscopy to measure intrinsic markers of orthorhombic symmetry. We identify that the film resolves this energetic conflict by switching its orientation by 90\textdegree~at an atomically-flat plane within its volume, not at the film--substrate interface. At either side of this ``switching plane'', characteristic orthorhombic distortions tend to zero to couple mismatched oxygen octahedra rotations. The resulting boundary is highly energetic, which makes it \textit{a priori} unlikely; by using second-principles atomistic modeling, we show how its formation requires structural relaxation of an entire film grown beyond a critical thickness measuring tens of unit cells. The switching plane breaks the inversion symmetry of the  \textit{Pnma} orthorhombic structure, and sharply joins two regions, a thin intermediate layer and the film bulk, that are held under different mechanical strain states. By therefore contacting two distinct phases of one compound that would never otherwise coexist, this alternative type of interface opens new avenues for nanoscale engineering of functional systems, such as a chemically-uniform but magnetically inhomogeneous heterostructure.

\end{abstract}

\section{}

Unlike conventional semiconductors, perovskite-structured transition metal oxides (TMO) offer broad possibilities for achieving functional electronic and magnetic properties, by exploiting the correlation between different degrees of freedom (spin, charge, lattice, orbital, topology)~\cite{Ngai2014,Tokura2017}. 
Epitaxial perovskite TMO thin-films and heterostructures further promote the coupling of degrees of freedom; on the structural level by setting the strain state via a careful selection of the substrate, as well as on the electronic level by charge transfer or screening effects. The effects of epitaxy have therefore been widely explored, leading to impressive modulation of physical properties, such as the tuning of the metal-to-insulator transition in nickelates~\cite{Catalano_2014}, or of the ferroelectric critical temperature in ferroelectrics~\cite{Haeni2004}. The success of this type of approach depends on both the sensitivity of the transition metal crystal field to any size modification of the oxygen octahedra, and on the coupling of the transition metal cations via the oxygen anions~\cite{Khomskii_2014}.

\begin{figure}[H]
\centering

\includegraphics[width=0.8\columnwidth]{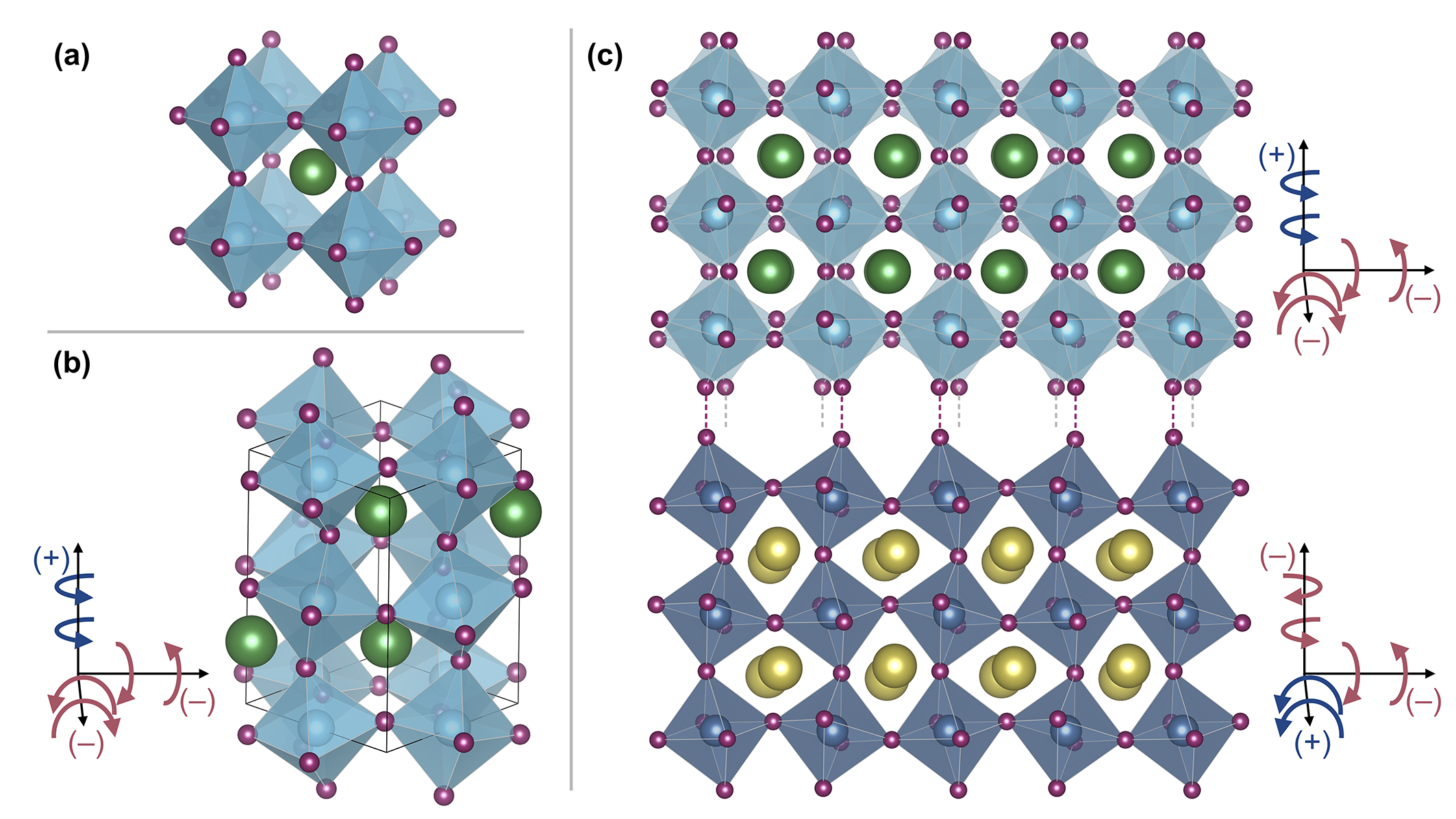}
\caption{Illustration of structural conflict at the rotation-coupled interface. Panel (a) shows the cubic perovskite unit cell, with six oxygen anions (purple spheres) forming an octahedron centered around each corner-sited $B$ cation (blue spheres). The $A$ cation (green sphere) sits in the middle of eight $B\text{O}_6$ octahedra. The orthorhombic unit cell variant of the perovskite structure is shown in (b), with the distortions and related out-of-phase $(-)$ and in-phase $(+)$ $B\text{O}_6$ rotation axes of its \textit{Pnma} symmetry. Panel (c) depicts the meeting of two \textit{Pnma} structures, the lower one having the in-phase OOR axis in the substrate plane, while it is perpendicular to the substrate plane in the upper one. By looking along a projection parallel to the in-phase OOR axis of the lower structure, we see that, at the interface of the two, only half of the apical oxygens of the upper structure can match directly to those of the lower structure (purple dashed lines). The remaining apical oxygens cannot form a match across the interface (grey dashed lines). Structural models prepared with the aid of VESTA~\cite{Momma2011}.}
\label{fig1}
\end{figure}

This coupling is founded on the $AB$O$_3$ perovskite structure, which can be described by a (pseudo-)cubic unit cell of corner-sited transition metal $B$-site cations that each combine with six oxygen anions to form eight corner-shared $B$O$_6$ octahedra surrounding a body-centered $A$-site cation;~\cite{Woodward1997} see Fig.~\ref{fig1}(a). For many TMO perovskites, the $A$-site cation is ``undersized'' for the space that it occupies. In this case, antiferrodistortive (AFD) tendencies drive the $B$O$_6$ oxygen octahedra to undergo rotations which shorten the $A$---O bonds while maintaining the $B$---O distances, thereby increasing energetic stability \cite{Buttner1992,Woodward1997,Zubko2011,Amisi2012,Miao2013}. The resultant set of oxygen octahedra rotations (OOR) is often described using the Glazer notation, which considers whether the octahedra rotate in-phase ($+$ sign) or out-of-phase ($-$ sign) along the three pseudocubic (``pc'') axes \cite{Glazer1972}. In the case of orthorhombic compounds---the most lattice system of TMO perovskites---the $B$O$_6$ octahedra rotate in-phase along one pc axis and out-of-phase along the other two; see Fig.~\ref{fig1}(b). Using the conventional \textit{Pnma} space group setting, the in-phase axis corresponds to the long orthorhombic axis $b_{\text{orth}}$ having lattice parameter $\approx 2 a_\text{pc}$ (where $a_\text{pc}$ is the lattice parameter of the pc unit cell), while $a_{\text{orth}} \approx c_{\text{orth}} \approx \sqrt{2} a_\text{pc}$. An extra component of energetic stabilization comes from the condensation of antipolar displacements of the $A$-site cations~\cite{Amisi2012,Miao2013,Woodward1997,Thomas1996,Benedek2013}. Deriving from other AFD instabilities, these are energetically-coupled to the OOR by a trilinear term~\cite{Amisi2012,Miao2013}. 
The more significant is the $X_5^-$ mode, involving positive and negative displacements of $A$-site cations along the $a_{\text{orth}}$ axis when following a lattice vector parallel to $b_{\text{orth}}$; see Supporting Information (SI) Fig. S1.
This mode itself constitutes a unique signature of the orthorhombic $a^{-}b^{+}c^{-}$ OOR pattern.~\cite{Mundet2024}

A key parameter determining the electronic coupling of the transition metal cations is the $B$---O---$B$ angle.
Since this in turn derives from the in-phase rotations of the corner-connected $B$O$_6$ octahedra, the connectivity of OOR across heterostructure interfaces presents an effective route towards controlling the $B$---O---$B$ angle~\cite{rondinelli_may_freeland_2012}. Moreover, aberration-corrected scanning transmission electron microscopy (STEM) has revealed that the OOR of a substrate or buffer layer can create an imprint on the OOR of a thin film, with an out-of-plane extent that depends on the symmetries either side of the epitaxial interface (cubic/orthorhombic, orthorhombic/orthorhombic), and on the rotation amplitudes of the materials.~\cite{Aso2013,Aso2014}

Besides OOR connectivity, epitaxial strain state is the main driver determining orthorhombic film orientation. When grown under biaxial epitaxial tension, the macroscopic strain energy of the film is basically minimized by $b_{\text{orth}}$ orienting out-of-plane~\cite{Masset2020,Choquette2016,Rotella2012,Meley2018,Meley2019,Mundet2024}. This is also the in-phase OOR axis of the film, while both in-plane pc axes have out-of-phase OOR. If a \textit{Pnma} $(101)_{\text{orth}}$ orthorhombic substrate is selected, it instead has $b_{\text{orth}}$ in-plane. As shown schematically in Figure~\ref{fig1}(c)\nocite{Momma2011}, along substrate $b_{\text{orth}}$, there is consequently a mismatch of the substrate's in-phase OOR with the out-of-phase OOR of the tensile strain state-favored film. This mismatch sets up a structural--energetic conflict, with the interface favoring an alternative film orientation, as detailed in the next section. Here, we unveil how the system resolves this conflict by creating a coherent structural interface that we term the \textit{switching plane}. Formed within a chemically-uniform TMO film, this corresponds to an atomically-thin boundary where its \textit{Pnma} structure switches orientation by 90\textdegree, thereby breaking the inversion symmetry of the orthorhombic perovskite lattice. At the same time, through detailed STEM characterization of the system of \LVO{} deposited on $(101)_{\text{orth}}$ \DSO{}, complemented by innovative second-principles modeling, we develop deep insights into the nanoscale structural--energetics of the orthorhombic film growth, and how the final atomic structure of the film depends on film thickness. As described later, these findings open unique opportunities for the deterministic engineering of novel functional properties.

\section{Results and Discussion}

\subsection{Film Nanostructure Analyses}

\begin{figure}[H]
\centering
\includegraphics[width=\columnwidth]{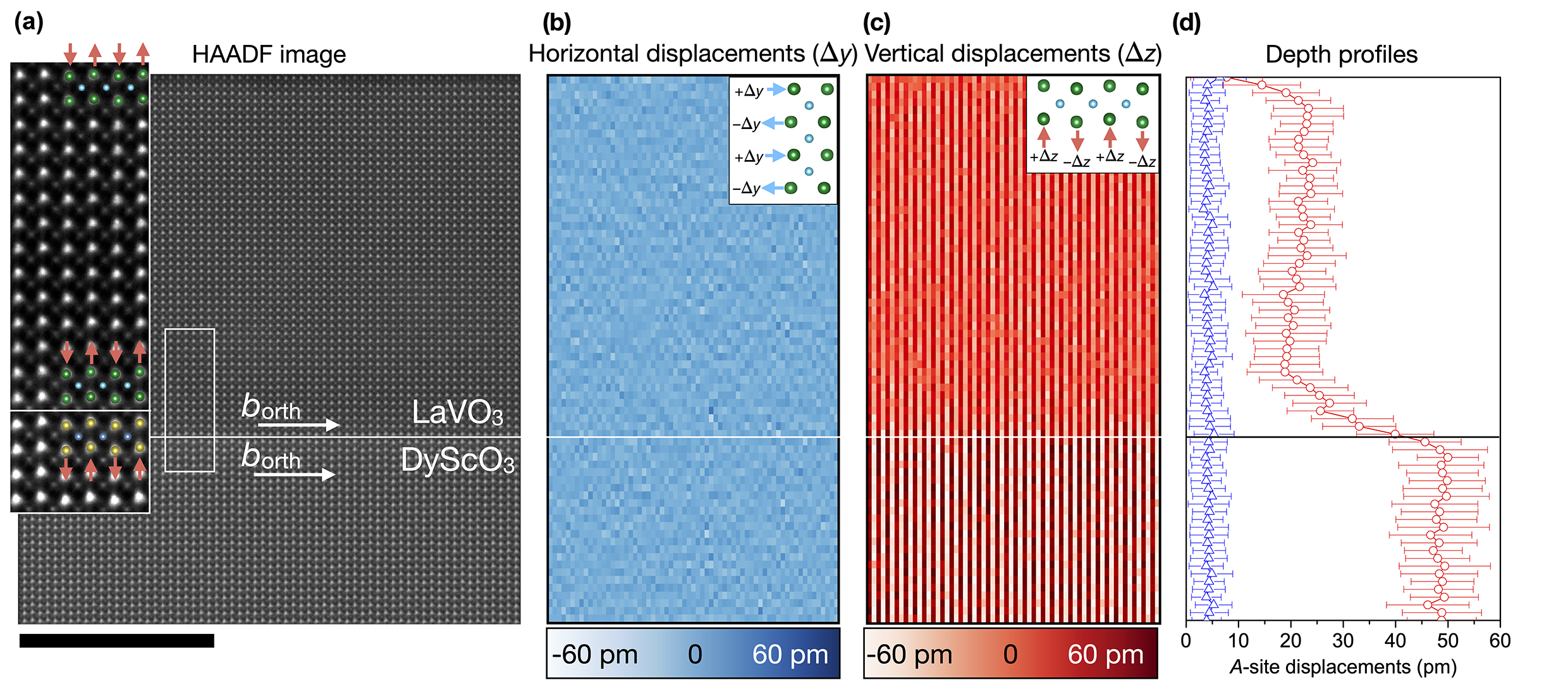}
\caption{STEM analysis of the 52~u.c.~\LVO{} on \DSO{} heterostructure viewed along the $[1\,0\,\Bar{1}]_{\text{orth}}$/$[1\,0\,0]_{\text{pc}}$ substrate zone axis. Panel (a) shows a HAADF image from the substrate up to the film surface, with the inset presenting a zoom of the region at the film--substrate interface indicated by the white rectangle. The OOR patterns of the film and substrate are revealed through the observation of the projected $X_5^-$ AM mode; the up--down displacement of successive Dy planes when moving along a left--right direction in the substrate propagates into the La sites across the whole film thickness. This qualitative impression is confirmed by a quantification of the $A$-site cation positions for (b) horizontal $\Delta y$ displacements and (c) vertical $\Delta z$ displacements. (Note that these two panels are compressed on the horizontal axis.) The average of the magnitude (i.e. modulus) of the $\Delta y$ and $\Delta z$ displacement values along a depth profile are shown in panel (d), with the curves color-coded according to panels (b) and (c). Only a vertical displacement of the $A$-site cations is seen, confirming that $b_{\text{orth}}$ remains in the substrate plane across the whole film thickness. In the film, there is a gentle decay in the magnitude of the AM over the first $\sim$8~u.c. from the interface, from the substrate value to that of the ``bulk film''. On this plot the error bars represent the measurement standard deviation. Scale bar: 10 nm.}
\label{fig2}
\end{figure}

To develop a basis for understanding switching plane formation, in Fig.~\ref{fig2} we first explore the consequence of OOR mismatch on a 52 pseudocubic unit cell (uc) thick \LVO{} film grown, by pulsed laser deposition, under ${\sim}0.5$\% biaxial tension on a $(101)_{\text{orth}}$ \DSO{} substrate (see SI Table S1). 
Fig.~\ref{fig2}(a) shows a STEM image of the sample, recorded on the $[100]_{\text{pc}}$/$[10\bar{1}]_{\text{orth}}$ zone axis of the substrate. Using the atomic number contrast of the high angle annular dark field (HAADF) detector, it depicts the brighter $A$-site and darker $B$-site cations. This enables imaging of the $X_5^-$ antipolar motion (AM) of the $A$-site cations (SI Fig. S1). 
Since, in the substrate, $b_{\text{orth}}$ lies in-plane, in Fig.~\ref{fig2}(a) its $X_5^-$ mode is seen in projection as a vertical AM of successive layers of $A$-site cations. We emphasize that, because of the trilinear energetic term coupling AM with the $a^{-}b^{+}c^{-}$ OOR, visualization of this mode can be used as a proxy for determining the in-phase OOR axis and relative amplitude~\cite{Zhang2013,Moon2014NC,Meley2018,Dominguez2020}. Fig.~\ref{fig2}(a) shows that the up--down AM propagates across the entire \LVO{} film. Evidently, the energetic cost of coupling the mismatched OOR of the strain-state favored film with those of the substrate is sufficiently high that the film has instead adopted the symmetry and $B\text{O}_6$ rotation configuration of the substrate, keeping $b_\text{orth}$ in-plane. As confirmation, quantified maps of the $A$-site cation positions in Fig.~\ref{fig2}(b) and (c) show that their projected AM remains purely in the out-of-plane pseudocubic $z$ direction across the film. The quantified line profile Fig.~\ref{fig2}(d) in turn shows that the magnitude of displacement decays over $\sim$8 uc from substrate into film, while the continuity of in-phase OOR from substrate into film is confirmed by annular bright field (ABF) STEM imaging around the film--substrate interface (SI Fig. S2).

\begin{figure}[H]
\centering
\includegraphics[width=0.4\columnwidth]{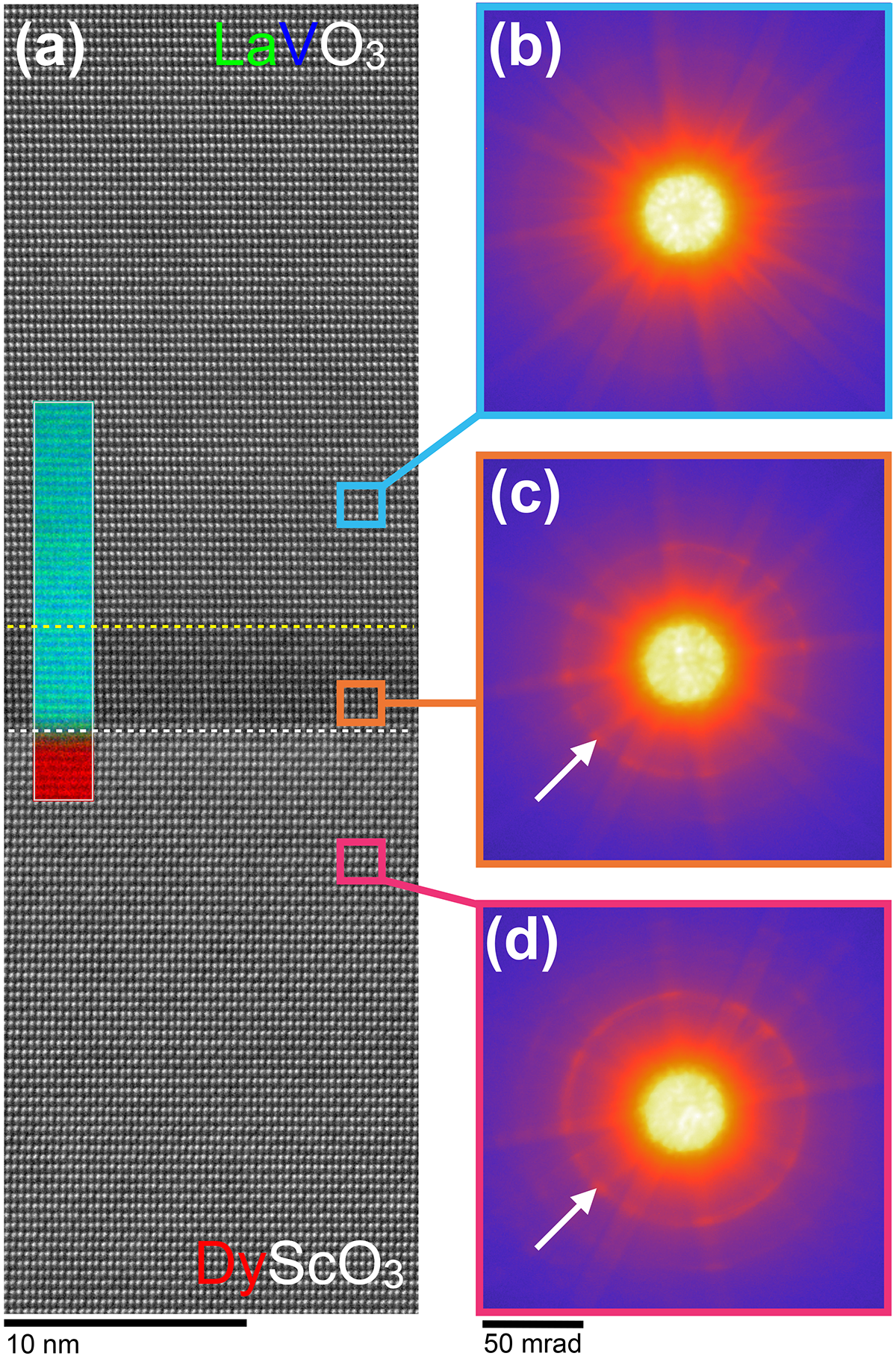}
\caption{(a) STEM-HAADF image of lower part of $\sim$110 uc \LVO{} film on \DSO{}, recorded on the $[110]_{\text{pc}}$ zone axis using a large inner collection semi-angle of $\sim$80~mrad. The white dashed line indicates the film--substrate interface. Remarkably, the first 10 uc of the film up to the yellow dashed line show a darker contrast than the bulk, even though the film composition is uniform, as illustrated by the overlaid EELS map for La (green), V (blue) and Dy (red). (b)--(d) PACBED patterns taken from the same film and zone axis at the analogous indicated positions in the substrate and film. The arrow in pattern (d) indicates a low angle first order Laue zone (FOLZ) ring, from the substrate having a $[11\Bar{1}]_{\text{orth}}$ zone axis. This FOLZ ring is maintained in pattern (c) that is taken from the region of film having a darker contrast. However, the ring disappears in pattern (b) for the film bulk, that in turn has a $[001]_{\text{orth}}$ zone axis.}
\label{fig3}
\end{figure}

Similar imposition of substrate symmetry and $B\text{O}_6$ rotations has been observed previously, in analogous $Pnma$/$Pnma$ film/substrate combinations, for films up to 35--40 uc thick~\cite{Proffit2008,Liao2016AFM,Yuan2018,Choquette2016,Mundet2024}.
Nevertheless, such a film has an increased macroscopic strain energy over one having $b_{\text{orth}}$ out-of-plane; a strain energy that scales with film thickness. Therefore, one may ask, if film growth is continued, is there a \textit{critical thickness} at which an epitaxial orientation transition will be induced? That is, a thickness at which the energetic gain from the film adopting $b_{\text{orth}}$ out-of-plane is enough to force the creation of an interface that couples the mismatched OOR of substrate and bulk film? As a first step in looking at this question, Fig.~\ref{fig3}(a) shows a lower magnification HAADF STEM image of a much thicker \LVO{} film. As established before~\cite{Meley2018}, the bulk lattice orientation of this film actually is determined by the biaxial tensile strain, with $b_{\text{orth}}$ out-of-plane. Structurally, therefore, the mismatched OOR of bulk film and substrate must have been accommodated. This image, recorded on the 
$[110]_{\text{pc}}$ ($[11\Bar{1}]_{\text{orth}}$ \DSO{}) zone axis, provides a first hint at this accommodation, from a band of darker contrast $\sim$10~uc thick that lies within the film, running along the film–substrate interface. Compositional analysis finds that this darker contrast is not chemical in origin (Fig.~\ref{fig3}(a), SI Figs. S3, S4). 
It is instead structural, as indicated by the accompanying position averaged convergent beam electron diffraction (PACBED) patterns of Figs.~\ref{fig2}(b)--(d).

\begin{figure}[H]
\centering
\includegraphics[width=\columnwidth]{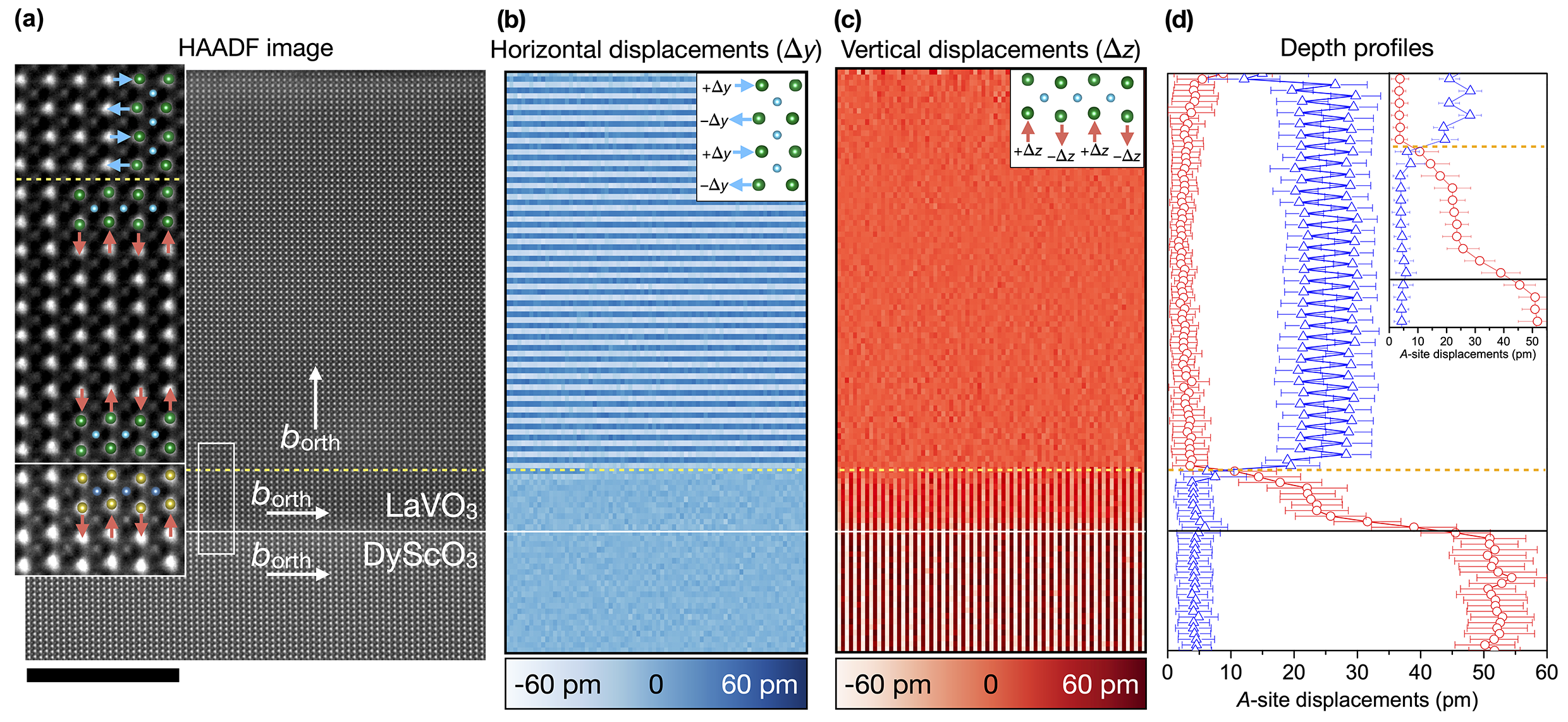}
\caption{STEM analysis of 81~uc~\LVO{} film grown on~\DSO{} viewed along the $[10\Bar{1}]_{\text{orth}}$/$[100]_{\text{pc}}$ substrate zone axis. (a) HAADF image across the full thickness of the film, with the inset showing a zoom of the region at the film--substrate interface indicated by the white rectangle. Based on this image, panels (b), (c) and (d) present quantified analyses of the $A$-site cation positions similar to those made in Figure~\ref{fig2}. From this data, it is seen that the substrate's $X_5^-$ AM mode propagates into the \LVO{} over $\sim$10~uc. At the 11th uc, the vertical AM mode abruptly decays to zero and then switches to a left-right displacement of successive La planes. This change in the orientation of the projected $X_5^-$ AM mode corresponds to a reorientation of the \textit{Pnma} unit cell to having $b_{\text{orth}}$ perpendicular to the substrate plane -- as favored by the biaxial strain state for the \LVO{} film. The plots (b), (c) (compressed on the horizontal axis) and (d) prove that this switch is abrupt, going from projected displacements that are purely vertical to purely horizontal at a well defined distance from the film--substrate interface, as indicated by the yellow dashed line. Indeed, the single uc shift in the position of this ``switching plane'' visible towards the left of the plot (b) mirrors a step edge in the substrate. The inset of plot (d) focuses on the region of film between the film--substrate and switching plane interfaces. In this intermediate layer, it is seen that the magnitude of AM decays rapidly from the substrate value to a plateau $\sim$5~uc in length, before decaying sharply to zero over the last few unit cells. On this plot the error bars represent the measurement standard deviation. Scale bar: 10 nm.}
\label{fig4}

\end{figure}

We now study this structural formation in more detail, in an 81 uc thick film. We use HAADF STEM imaging on the $[100]_{\text{pc}}$/$[10\bar{1}]_{\text{orth}}$ zone axis of the substrate to evaluate the $A$-site AM of the orthorhombic lattices. Compared to the alternative approach of ``direct'' imaging of in-phase $B\text{O}_6$ rotations using, for instance, ABF, this enables measurement of the orthorhombic distortions of $b_{\text{orth}}$ both in-plane and out-of-plane in a single image. It is further robust to residual aberrations and sample mis-tilts, such that quantitative atomic data can be taken across the entire $\sim$35~nm thickness of the film, with a 4k$\times$4k pixel resolution. Fig.~\ref{fig4}(a) presents an example image. In the \DSO{} substrate at the bottom of the zoomed inset, the $A$-site cations have an up--down AM for $b_{\text{orth}}$ in-plane. In contrast, higher up in the \LVO{} film they displace left--right, indicating that it has adopted $b_{\text{orth}}$ out-of-plane, as favored by tensile strain. Surprisingly, however, the transition between these two orientations does not occur at the film--substrate interface. Instead, the AM of the substrate propagates into the film over $\sim$10~uc, and only \textit{then} switches orientation. In Fig.~\ref{fig4}(b) and (c), this observation is studied with quantified maps of $A$-site left--right ($\Delta y$) and up--down ($\Delta z$) displacements. In a segment that we term the \textit{intermediate layer} (IL), the AM of the substrate continues into the \LVO{}, before a sharp segue to the left--right AM of the film bulk. Fig.~\ref{fig4}(d) shows depth profiles of the average $A$-site displacements across the IL. Going from the substrate into the thin film, the projected AM displacements remain purely parallel to the pseudocubic $z$-axis. At the same time, their amplitude decays over a few unit cells to a plateau of  $\sim$5~uc length. At the end of this, the amplitude further decays rapidly, reaching a value near zero at the top of the IL. At this point -- which we term the switching plane -- projected AM displacements switch sharply to being parallel to the pseudocubic $y$ axis, corresponding to an out-of-plane $b_{\text{orth}}$. The displacements then increase in amplitude over a few uc to reach their final value for the bulk film structure. The amplitude of 
AM in the IL plateau region is approximately equal to that of the film's bulk structure. In agreement with this analysis, ABF STEM along substrate $[010]_\text{orth}$ shows directly how the in-phase OOR of the IL tend to zero at the switching plane (SI Fig. S5).

The initial continuity of substrate symmetry into the film over some 10 uc, followed by a switch to $b_\text{orth}$ out-of-plane, explains the PACBED results along $[110]_\text{pc}$ in Fig.~\ref{fig3}. The corresponding $[11\bar{1}]_{\text{orth}}$ zone axis of the \DSO{} substrate gives it a strong first order Laue zone (FOLZ) ring at a low scattering angle of $\sim$60~mrad, owing to a symmetry distance doubling along the beam path direction~\cite{Nord2019}. Because of the structural continuity, this FOLZ ring remains in the IL; however, it disappears in the film bulk when the zone axis \textit{switches} to $[001]_{\text{orth}}$. 4D-STEM proves that this disappearance occurs discretely, when stepping one uc across the switching plane (SI Fig. S6). 
Together with simulations using $\mu$STEM~\cite{Forbes2010, Allen2015}, the PACBED analysis also explains the darker contrast from the IL in Fig.~\ref{fig3}(a). Specifically, this arises from strong elastic scattering into the FOLZ rings of the substrate and IL that consequently leads to them having anomalously low thermal diffuse scattering and hence decreased signal on the HAADF detector (SI Figs. S7, S8). 
As a result, the dark contrast along the $[110]_\text{pc}$ zone axis itself constitutes an indication of IL/switching plane formation.

To summarize, thicker \LVO{} films adopt the orientation expected from the substrate-imposed tensile strain, \textit{except} for an initial IL which keeps the substrate orientation. The switch between the two orientations occurs sharply, at the switching plane. As described later, the local atomic topography at the switching plane enables a subtle transition between the mismatched OOR of the bulk film and substrate orientations. Associated with this, incompatible orthorhombic distortions tend to a value of zero, while compatible ones propagate freely. Since it contains strong local distortions compared to the $Pnma$ structure that forms the basis of the \LVO{} lattice, this novel interface is a highly energetic boundary.

Our findings have various implications. First, the critical thickness for transitioning to the strain-state determined bulk film orientation, tied to switching plane and IL formation, is clearly between 52 and 81 uc. By using X-ray diffraction to monitor the appearance of a half-order reflection for $b_\text{orth}$ out-of-plane in a film thickness series, this value is further narrowed to between 60 and 74 uc (SI Fig. S9). 
Second, given that, during deposition, the film initially grows in structural continuity with the substrate, its atomic lattice must dynamically restructure after reaching the critical thickness, when the bulk film switches its orientation. While the required energy for restructuring is presumed to be thermal from the substrate heating, the mechanism remains an open question. Finally, we hypothesize that the switching plane combined with IL represent an energetic minimum, as compared to forming an OOR-coupled interface that coincides with the film--substrate interface. This reminds of the stand-off effect of misfit dislocations, where the origin of the strain-releasing defects sits in the elastically weaker material, a few uc far from the layer--substrate interface  \cite{Mader1991,Gutkin1994,Lu2019}. Given its importance, we now explore the energetics of the system \textit{via} second-principles calculations.

\subsection{Second-Principles Calculations}

Making appropriate simulations is a non-trivial task, owing to the need to include tens of uc thickness in the film. The ``standard'' approach of density functional theory calculations, as for instance used to calculate the biaxial strain determined \LVO{} structure~\cite{Meley2018}, is therefore unviable. In order to address this challenge, we innovate a second-principles modeling approach~\cite{Schmitt2020th,Ghosez22}. 

Building second-principles effective atomic potentials remains very challenging and a substantial work (both for computing an extensive training set of first-principles data and for fully validating the model), which explains why only few high-quality models are presently available. For our simulations, we turn to a system of \CTO{}, that we adapt to incorporate physics and constraints analogous to our experimental system. In doing so, we leverage a model that has been previously validated and importantly demonstrated to be accurate for reproducing inhomogeneous structures such as twin walls; see Schmitt et al.~\cite{Schmitt2020th}, Zhang et al.~\cite{Zhang2023} and SI Fig. S10. 
Moreover, \CTO{} is the prototypical \textit{Pnma} perovskite, such that our approach implicitly demonstrates the generality of the concept underpinning switching plane formation, that depends on applied constraints rather than specific compound. Finally, \CTO{} is non-magnetic, aiding tractability of the calculations.

\begin{figure}[H]
\centering
\includegraphics[width=0.5\columnwidth]{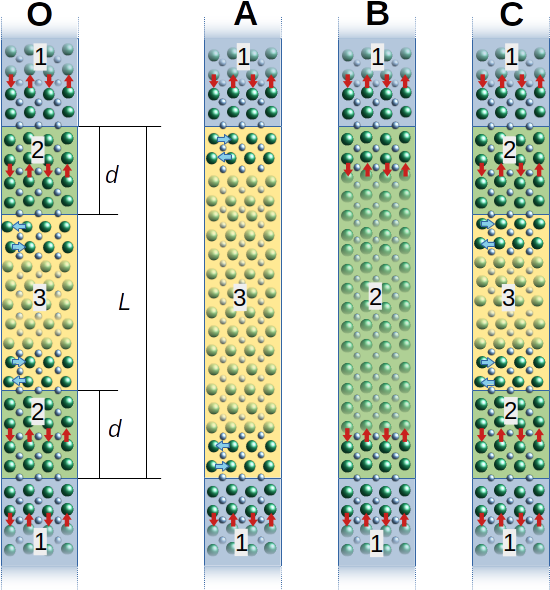}
\caption{Structural simulations. (O) Illustration of the supercell layout used for the second principle simulations. The supercells are composed of three parts. Segment 1 mimics the substrate, having $b_{\text{orth}}$ in-plane but with exaggerated OOR and AM. Its atomic positions are fixed during the relaxation. Segments 2 and 3 mimic the film, with segment 2 having $b_{\text{orth}}$ in-plane, compared to it being out-of-plane for segment 3. During the simulation, their atomic positions are free to relax. We denote $L$ for the total number of~uc layers in the film and $d$ for the number of~uc layers in segment 2. Periodic boundary conditions apply. Three distinct configurations for the film are possible: A) the whole film consists of $b_{\text{orth}}$ out-of-plane; B) the whole film consists of $b_{\text{orth}}$ in-plane; C) consists of a mix of both orientations with various numbers of uc layers $d$ in segment 2.}
\label{fig5}

\end{figure}

The atomic structure ``O'' in Fig.~\ref{fig5} defines the basic supercell setup for the simulations. It makes use of periodic boundary conditions that duplicate the interface. Segment 1 mimics the atomic environment created by the substrate: it is a region of \textit{Pnma} structure with $b_{\text{orth}}$ in-plane, in which the orthorhombic distortions (OOR and $A$-site AM) are artificially-amplified and frozen to replicate the distortion mismatch and biaxial strain imposed by the substrate at the~\LVO{}--\DSO{} interface. Segments 2 and 3 comprise the thin film. To represent the IL, segment 2 has a \textit{Pnma} structure with $b_{\text{orth}}$ in-plane, while segment 3 conforms to the $b_\text{orth}$ out-of-plane structure. Simulations are made for a series of values of absolute film thickness $L$, as specified in uc layers. For each $L$, we consider the energies of the three configuration types illustrated in Fig.~\ref{fig5}: A) film of energy $E_\text{A}(L)$ that consists only of $b_\text{orth}$ out-of-plane ($d=0$); B) film of energy $E_\text{B}(L)$ that consists only of $b_\text{orth}$ in-plane ($d=L/2$); C) films of energies $E_\text{C}(L,d)$ containing a mix of both orientations with IL thickness $d$ ($0<d<L/2$). Each structure in the sequence of $0\le d \le L/2$ is allowed to relax (atomic positions in segments 2 and 3 and cell parameter out-of-plane), after which their energetic values are compared.

\begin{figure}[H]
\centering
\includegraphics[width=\columnwidth]{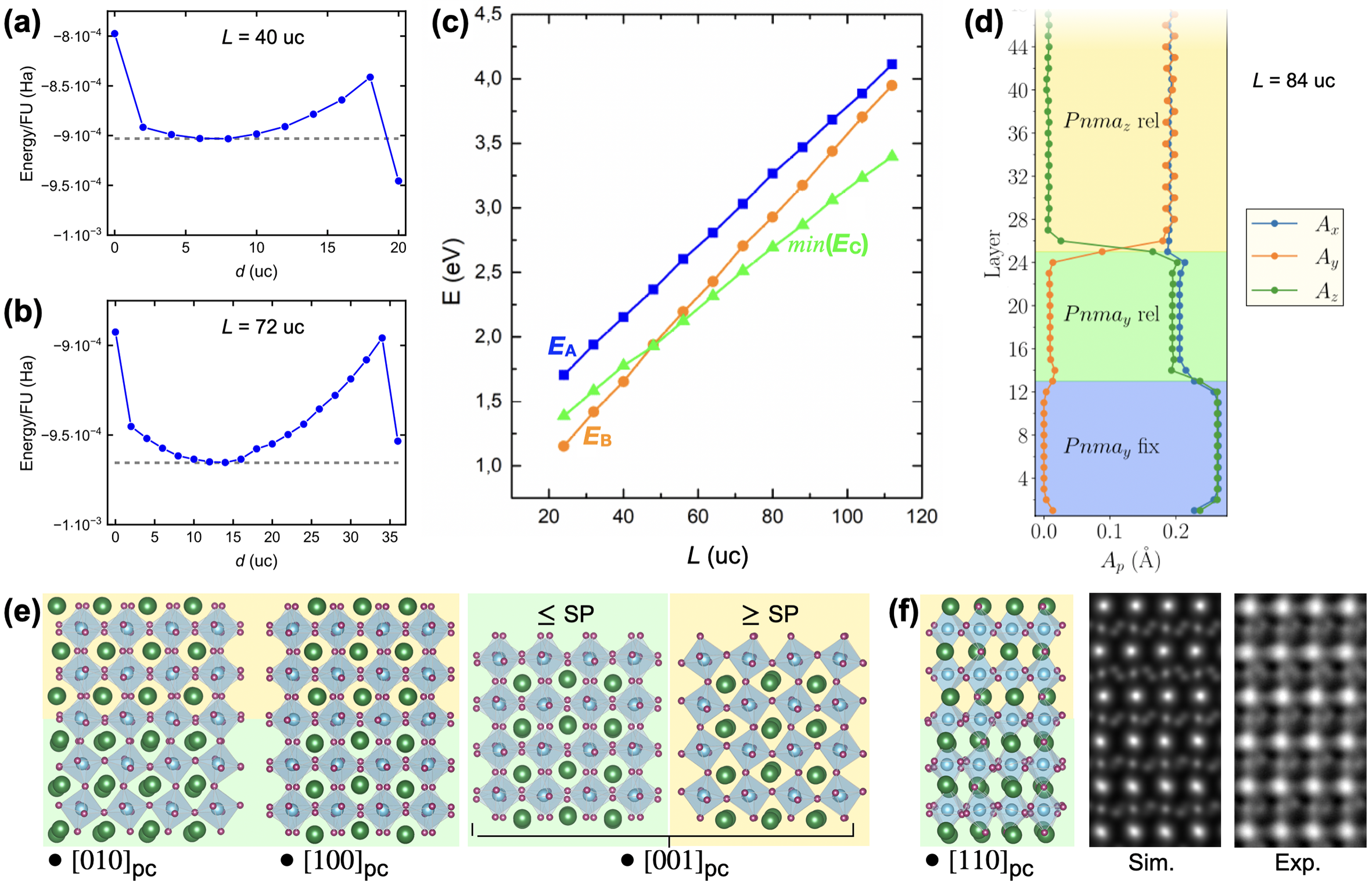}
\caption{Structure energy as a function of IL thickness $d$ for supercells of dimension (a) $L=40$~uc and (b) $L=72$~uc. To help compare values, the grey dashed lines indicate the minimum energy for films having configuration C. (c) Relative energies $E_\text{A}$, $E_\text{B}$ and $min(E_\text{C})$ of configurations A, B and C against $L$. The curve $min(E_\text{C})$ represents the minimum possible energy of configuration C, by appropriately optimizing $d$ for each $L$. (d) Depth profile of the amplitude of AM displacements of $A$-site cations along the three pseudocubic axes across the minimum energy supercell having $L = 84$~uc. Because of the symmetric boundary conditions only the lower half of the supercell is shown. (e) Projections of the $L=84$~uc minimum energy configuration across the switching plane along different axes, with backgrounds color-coded according to Fig.~\ref{fig5}; the switching plane is at the junction of the green and yellow sections. (f) Projection along the $[110]_\text{pc}$ axis with the upper ``bulk film'' part having a $[001]_\text{orth}$ zone axis. From this, an inverted ABF STEM image is simulated (substituting Ca and Ti by La and V respectively) and compared to an averaged experimental image from a sample of $\sim$110 uc~\LVO{} grown on~\DSO{}.}
\label{fig6}

\end{figure}

From the simulation results, Fig.~\ref{fig6}(a) presents the energetic evolution of a relatively thin film ($L=40$), in function of $d$. The highest energy is observed for $d=0$, which corresponds to placing the switching plane at the film--substrate interface (configuration A). As $d$ increases, the energy decreases to a minimum at $d=8$, but then rises again. Finally, for $d=L/2$, the curve shows a sharp decrease to its lowest energy, making B the most favorable configuration; \textit{i.e.}, the symmetry-imposed structure, as for \LVO{} films under critical thickness. Considering the results for a thicker film with $L=72$ in Fig.~\ref{fig6}(b), a similar curve is seen. However, there is one key difference: the minimum energy in the initial concave (at $d=14$) is now lower than that of the final, sharp minimum at $d=L/2$. Therefore, configuration C having an IL and implicit switching plane is now the most energetically favorable. By repeating the calculations for $24 \le L \le 104$ (SI Fig. S11), 
the summarizing plot in Fig.~\ref{fig6}(c) is determined, in which $min(E_\text{C})$ corresponds to the minimum $E_\text{C}(L,d)$ by adjusting $d$ to an optimized value $d_{min}$ for each $L$. Remarkably, the second-principles simulations reproduce the experimentally-observed transition, since the lowest energy configuration changes from configuration B with $b_{\text{orth}}$ uniformly in-plane to the mixed $b_{\text{orth}}$ in-plane/out-of-plane configuration C at a critical film thickness of, in this case, $L\approx48$ uc. 

\subsection{Structural--Energetics Analysis}

To obtain physical insight into these results, it is instructive to decompose the energies as follows:
\begin{eqnarray}
E_\text{A} (L) &=& E_1 + L E_3 + 2 E_{13} \\
E_\text{B} (L) &=& E_1 + L E_2 + 2 E_{12} \\
E_\text{C} (L,d) &=& E_1 + 2d E_2 + (L-2d)E_3 + 2 E_{12} + 2 E_{23}
\end{eqnarray}
in which $E_1$ is the energy of the substrate (segment 1), $E_2$ and $E_3$ are the energies per u.c. of segments 2 and 3, and $E_{ij}$ are the interface energies between segments $i$ and $j$.  

In Fig.~\ref{fig6}(c), the energy $E_{\text{A}}(L)$ for the $b_{\text{orth}}$ out-of-plane structure (blue circles) is higher than energy $E_{\text{B}}(L)$ for the $b_{\text{orth}}$ in-plane structure (orange upwards-pointing triangles) at small thicknesses. This result is driven by the interface energy, with the film preserving the orientation of the substrate to ensure better continuity of the atomic distortions (i.e. $E_{13} > E_{23}$). However, in terms of elastic energy, the out-of-plane structure is favored (i.e. $E_3$ is lower than $E_2$). This means that both curves have a different slope. Extrapolation implies that they will cross at $L \approx 140$ uc, when sufficient thickness is achieved for the cost of interface energy to be compensated by the elastic energy gain. We further notice that the curves are almost linear, which means that $E_{13}$ and $E_{12}$ are almost independent of $L$. 

While a crossing from configuration B to A could be in line with usual expectations, our second-principles simulations point out that such a transition will never actually happen. Instead, in agreement with our experimental findings, the film will prefer to switch to the mixed configuration C (green downwards-pointing triangles) -- even though the latter creates an additional energy cost $E_{23}$ for the switching plane interface. At small thicknesses, $E_\text{B}(L)<E_\text{C}(L)<E_\text{A}(L)$, highlighting that $E_{12}+E_{23}<E_{13}$. This confirms that it is energetically more favorable to form the OOR-coupled interface within the film, rather than directly at the film--substrate interface. Nevertheless, this alone is insufficient to explain the observed behavior in both simulations and experiments. If $E_{23}$ is assumed to be constant, then the slope of $E_\text{C}(L)$ would be the same as that of $E_\text{A}(L)$ and also the system would keep $d$ minimum (i.e. $d=1$) at any $L$ (since $E_2 > E_3$). In contrast, the slope of $E_\text{C}(L)$ is smaller than that of $E_\text{A}(L)$ and slightly decreasing with $L$, and both experimentally and numerically $d_{min}\gg1$. For, crucially, the switching plane energy $E_{23}$ depends on $d$ and $L$, and progressively decreases with them. Physically, this is because providing larger thicknesses of film on both sides of the switching plane decreases $E_{23}(L,d)$, by better accommodating the disparity of atomic distortions. So, on the one hand, the switching plane wants to move away from the film--substrate interface (and surface) to the interior of the film to decrease $E_{23}(L,d)$ while, on the other hand, the film wants to keep $d$ as small as possible to minimize the elastic energy. The transition from configuration B to configuration C appears at a critical thickness of the film, at which $d$ is large enough for the cost of $E_{23}(L,d)$ to become small, while $(L-d)$ is also sufficiently large for the lowering of elastic energy to compensate the switching plane formation.

In Fig.~\ref{fig6}(d), depth profiles are plotted of the $A$-site AM amplitudes for the $min(E_\text{C})$ configuration of an example film above the critical thickness. $A_y$ and $A_z$ are respectively equivalent to the $\delta_y$ and $\delta_z$ values measured along the $[100]_\text{pc}$ zone axis in Figs.~\ref{fig2} and~\ref{fig4}. At the film--substrate interface, the amplitude $A_z$ of the substrate AM propagates into the film, but decays over 2--3~uc to a 10 uc plateau. At the end of the plateau, $A_z$ sharply decays towards zero amplitude at the switching plane. At that point, the amplitude of $A_y$ sharply increases from zero to that of the bulk thin film structure of segment 3. These AM transitions mirror those measured for the~\LVO{}/\DSO{} system in Fig.~\ref{fig4}, vindicating the similarity of model to experiments. Notably, the AM amplitude modulation across the switching plane occurs much more sharply than that at the film--substrate interface in films grown below the critical thickness, where the substrate symmetry is preserved (see Fig.~\ref{fig2}). This can be understood as the system confining the distorted and highly energetic structure around the switching plane to a small volume for energetic minimization. Not only is this discrete switch between two structural phases experimentally confirmed by the STEM and PACBED measurements of \LVO{} on $(101)_\text{orth}$ \DSO{}, but it represents a case distinct from an alternative type of energetic conflict that was set up in a system of La$_{2/3}$Ca$_{1/3}$MnO$_3$ grown on $(12\bar{1})_{\text{orth}}$ NdGaO$_3$, where a smooth modulation of orthorhombic distortions over $\sim$14 uc was observed.~\cite{Zhang2022}

Interrogation of the structural model, and its comparison to STEM data along different zone axes, allow us to elucidate further the nature of atomic structure transition across the switching plane. Fig.~\ref{fig6}(e) shows projections of the structural model on the three pseudocubic axes. $[010]_\text{pc}$ is projected along the $b_\text{orth}$ axis of the substrate, and illustrates how the mismatched OOR of bulk film and substrate interface by a sharp flattening or damping within 2 uc either side of the switching plane. Along $[100]_\text{pc}$, as well as AM we visualize out-of-phase OOR modes for both substrate and bulk film orientations. These two modes do not face the same connectivity problem and, because of this compatibility, we see that the mode propagates undamped across the switching plane. Experimental ABF STEM data are consistent with this finding; see SI Fig. S12. 
In order to study OOR connectivity on the out-of-plane $[001]_\text{pc}$ axis, we present two projected slices, each 2 uc thick: one that includes the switching plane uc and the uc below ($\leq$ SP) and the other including the switching plane uc and the uc above ($\geq$ SP). It is evident that, on this axis, the OOR mode switches from out-of-phase to in-phase very sharply, with negligible damping. Such an abrupt switching in OOR modes along the out-of-plane axis has previously been seen in rhombohedral/orthorhombic heterostructures~\cite{He2015}, and occurs because the apical oxygen effectively acts as a free pivot point for out-of-plane rotations. In a cross-section sample, these rotations can be monitored by measuring the symmetry of O dumbbells or clusters imaged on a $[110]_\text{pc}$ axis~\cite{He2015}. Fig.~\ref{fig6}(f) shows the appropriate projection, where the bulk film has a $[001]_\text{orth}$ zone axis. Above the switching plane, the O dumbbells are mirrored by a horizontal plane -- indicative of in-phase OOR along the out-of-plane axis. Below it, as the OOR go to out-of-phase, the O clusters lose this mirror symmetry. From the model, an (inverted) ABF image is simulated (``Sim.''), substituting the Ca and Ti cations by La and V respectively, in order to be more easily compared to the experimental data. The experimental image (``Exp.'') on the right bears a resemblance to the simulation, supporting the simulation-based hypothesis that out-of-plane OOR switch directly from out-of-phase to in-phase across the switching plane.

\section{Summary and Perspectives for Functional Property Engineering}

In summary, through experiments and simulations, we have revealed a complex ``phase space'' for guiding the design choice of orthorhombic film growth, set by factors of epitaxial strain, OOR connectivity and film thickness. As shown by simulations in Fig.~\ref{fig6}, under our chosen $Pnma$/$Pnma$ parameters, films grown below a critical thickness have a structural phase set by OOR connectivity. Above the critical thickness, the film instead adopts a two phase structure: a bulk structure set by epitaxial strain, and a thin IL adjacent to the substrate which instead follows the substrate orientation. The calculations further show that considering only the epitaxial strain in OOR/strain imposed systems~\cite{Meley2018,Zhang2022} gives the highest film energy of all structural variants, and so does not correctly predict the film's complete atomic structure. Indeed, our simulation approach could also be useful for modeling other systems, such as manganite thin films grown under alternative energetic impositions~\cite{Zhang2022}, or whose energetics remain unexplained~\cite{Yang2021}.

In films grown beyond the critical thickness, the switching plane is formed between the two phases of the film. Its formation is driven solely by energetics, with no stochastic role of grain or domain nucleation. One consequence is that it appropriates the atomic flatness of the substrate, as for instance seen by the switching plane mirroring the substrate step edge in Fig.~\ref{fig4}. Simulations nevertheless imply that the IL thickness $d$ derives from a broad energetic minimum (see energetic curves in Fig.~\ref{fig6}b and SI Fig. S11.) 
The exact IL thickness is therefore expected to be sensitive to subtle factors during switching plane formation. This could explain small differences in IL thickness that are sometimes observed at different sampling points for a single deposited film, and is an aspect we are investigating with further experiments. While the atomic landscape of the IL and switching plane is confined to a relatively short length-scale of some 10--15 uc, it is a consequence of energetics acting over the entire film, grown beyond a significant critical thickness of tens of uc. Indeed, its formation can only be predicted by including the film's full atomic structure in a simulation. Because of this subtlety, conceivably its presence has been missed in previous work~\cite{Kan2013}.

In TMO materials, it is widely known that heterostructure interfaces that break the bulk lattice symmetry can be exploited to create functional properties beyond the scope offered by the unbound crystal~\cite{Hwang2012}, such as the formation of a two-dimensional electron system at the interface between insulating LaAlO$_3$ and SrTiO$_3$ compounds~\cite{Ohtomo2004}. Within single-phase compounds, crystalline boundaries such as domain walls in ferroelectric or ferroelastic materials have themselves produced emergent properties distinct from their bulk counterparts~\cite{Tagantsev2010,Langenberg2019,Farokhipoor2014}. In comparison to these priors, the 90\textdegree~orthorhombic structure rotation at the switching plane breaks the inversion symmetry of the lattice, which could lead to the emergence of topological states~\cite{Lesne2023}. Further, the switching plane separates two regions of the same chemical compound (IL and film bulk) that are held under distinct mechanical boundary conditions. Because it is induced by simple opposition of energetic influences on epitaxial film growth, it is in principle applicable to other \textit{Pnma}/\textit{Pnma} film/substrate combinations. Given that the orthorhombic lattice is the most common perovskite structure \cite{Glazer1972}, this gives a wide potential for creating the IL and switching plane in other compounds. Two interesting candidates are manganite or nickelate films. In these, we can exploit the different strain conditions of the IL and bulk film to engineer adjacent regions that, despite being of the same compound, have distinct electronic and magnetic properties, with the switching plane forming an unprecedentedly sharp interface between them. The IL of such a structure could, for instance, be used to pin the magnetic phase in a \LMO{} film grown on \GSO{} to the $b_{\text{orth}}$ reorientation, thus forming a chemically-uniform material that is magnetically inhomogeneous \cite{Schmitt2020}. Other possibilities for engineering novel functional properties also open up, such as leveraging this atomically-flat interface to create a 2-dimensional conductor~\cite{Catalan2012}. Finally, the control of the crystallographic orientation of the orthorhombic layer provided by the strong interfacial coupling of the OOR, that dominates over the lattice strain up to a critical thickness, offers a way to engineer complex heterostructures where the functional properties, for instance the magnetization axis \cite{Marshall1999,Fujita2024}, can be set by appropriate choice of the substrate surface plane.
As such, the switching plane/IL formation and defined phase space for orthorhombic film growth provide new strategies towards the deterministic engineering of functional properties at the nanoscale.



\section{Methods}
\subsection{Thin Film Growth}

The \LVO{} films are grown on $(101)_\text{orth}$ \DSO{} substrates by pulsed laser deposition using an excimer KrF laser run at 1 Hz repetition rate and at high pulse fluence (2 J/cm$^2$). Deposition occurs on substrates heated from 800$^{\circ}$~C to 900$^{\circ}$~C (standard value of $\sim$880$^{\circ}$~C) in a $5 \times 10^{-7}$~mbar oxygen atmosphere from a ceramic target of LaVO$_4$; cooling is performed under the same oxygen pressure. \textit{In situ} reflection high energy electron diffraction reveals that the deposition evolves from a layer-by-layer growth mode during the first few unit cells to a mainly step-flow mode. Atomic force microscopy identifies the high surface quality of the films: the film topography for all the thicknesses displays a step-and-terrace structure, mirroring the $(1\,0\,1)_{\text{orth}}$ \DSO{} substrate surface. For more details, see reference~\cite{Meley2019}.\\

\subsection{X-Ray Diffraction}

The scans were acquired with a X’Pert PRO PanAlytical diffractometer equipped with a Ge(220) monochromator and a triple-axis analyzer. \\

\subsection{Scanning Transmission Electron Microscopy}

Samples for STEM were prepared either by a combination of mechanical polishing using an Allied High Tech MultiTech polishing system, followed by argon ion beam milling with a Gatan PIPS II system to electron transparency, or by focused ion beam milling using a Zeiss NVision 40. 
All STEM data were acquired using a monochromated, double aberration-corrected FEI Titan Themis 60-300 operated at a high tension of 300 kV and using a probe semi-angle of convergence of 20.7 mrad and beam current of $\sim$40 pA. HAADF STEM images were acquired using a Fischione photomultiplier tube (PMT) detector. Unless otherwise stated, detector collection semi-angles of $\sim$50--200 mrad were applied, using a nominal camera length of 115 mm. For quantitative analysis of atomic column positions, image stacks were recorded at 90$^{\circ}$ rotations between consecutive frames. These stacks then underwent rigid and non-rigid alignment using the SmartAlign software~\cite{Jones2015}, in order to reduce artefacts from system noise and scan drift. In order to decouple scan distortions from the atomic column row directions, the images were acquired with an angle of $\sim$10--15$^{\circ}$ between the fast scan direction and one of the principal directions of the atomic rows. This methodology was successfully applied for frames up to 4k $\times$ 4k pixels in size. In the HAADF images, $A$ and $B$ cation positions were identified by fitting two-dimensional Gaussian functions using Atomap~\cite{Nord2017}. From these measurements, quantified maps and averaged depth profiles of $\Delta y$ and $\Delta z$ AM displacements were calculated using custom Python scripts.

When ABF images were recorded to visualize O sites, these were acquired simultaneously to the HAADF image series with the same nominal 115 mm camera length, using a Gatan PMT detector mounted at the entrance of a Gatan GIF Quantum ERS, giving collection semi-angles of $\sim$10.6--24.3 mrad. Scan distortions were corrected by first performing rigid and non-rigid alignment of the HAADF image stack, and then applying the HAADF-determined corrections to the simultaneously-acquired ABF image stack. In this way, the rigid and non-rigid alignment is unaffected by distortions from residual aberrations, sample mis-tilts or improper defocus that can have a stronger effect on the phase-contrast ABF image than incoherent HAADF image. O column positions were similarly identified using Atomap, with $B$---O---$B$ angle plots calculated using a custom Python script. Sub-frame averaging was performed with the SmartAlign Template Matching Module.

Analytical data were acquired using the same instrument, with STEM-EELS maps recorded with the Gatan GIF Quantum ERS, and energy-dispersive X-ray spectroscopy (EDXS) data recorded using the FEI/Thermo Fisher Scientific ChemiSTEM 4 quadrant silicon drift detector system. The EDXS data were recorded using the same probe semi-angle of convergence and HAADF detector set up as previously described. In contrast, the EELS data were recorded using a monochromated setup, with a probe semi-angle of convergence of $\sim$18 mrad, spectrometer semi-angle of collection of $\sim$36 mrad, and HAADF angles of $\sim$74–170 mrad. The STEM-EELS elemental maps were prepared using the analysis functions in Gatan DigitalMicrograph 3.5. Note that the V map of Fig.~\ref{fig3} and SI Fig. S3 
was integrated only from the V $L_{3,2}$ peaks, in order to avoid contribution from the O $K$-edge. STEM-EDXS chemical maps and line profiles were prepared using Thermo Fisher Scientific Velox 3.10 software, applying a Schreiber-Wims ionization cross-section model for elemental quantification.

PACBED patterns were acquired in two different ways. The PACBED data presented in the main text were acquired using the Gatan Ultrascan CCD camera of the GIF. The PACBED data presented in the SI were acquired using a MerlinEM (Quantum Detectors) Medipix3. For this latter measurement, the microscope was operated at 200 kV high tension.

STEM image simulations were performed with Dr. Probe software~\cite{Barthel2018}, using the imaging conditions detailed above for experimental acquisition, with aberrations and defocus set to 0. A source size of 0.015 nm was applied. PACBED simulations were performed using $\mu$STEM~\cite{Allen2015}.\\

\subsection{Second-Principle Calculations}

A second-principles model of \CTO{} was used to perform structural relaxations with applied constraints, as described in the main text, mimicking the conditions exhibited by the \LVO{} thin film grown on the \DSO{} substrate. Relaxations were performed using supercell sizes up to $124 \times 4 \times 4$ repetition of the 5-atom pseudocubic perovskite structure (9920 atoms). The atomic relaxations were performed using the \textsc{Multibinit} package~\cite{GONZE2020}, until the maximum force become smaller than $10^{-4}$ $\si{\electronvolt\per\angstrom}$.

The second-principles model was built with the \textsc{Multibinit} package which implements the second-principles approach outlined in references~\cite{Wojdel2013} and~\cite{Escorihuela-Sayalero2017}. This method relies on a Taylor expansion of the potential energy surface (PES) around the reference cubic structure in terms of all structural degrees of freedom, with coefficients then determined from first-principles data. In this scheme, the energy includes harmonic and anharmonic contributions in terms of individual atomic displacements, macroscopic strains and their couplings, with the long-range dipole-dipole interaction treated explicitly. At the harmonic level, the coefficients are exactly those directly computed from density functional perturbation theory. At the anharmonic level, the most relevant terms are selected and their coefficients are fitted in order to reproduce the energies, forces and stresses computed from density functional theory (DFT) for a set of configurations properly sampling the PES.

The training set of first-principles DFT data contained more than 5000 structures, calculated with the \textsc{Abinit} software package, making used of a plane-wave pseudopotential approach~\cite{Gonze2002,Gonze2009,Gonze2016,GONZE2020}. The DFT calculations were performed within the generalized gradient approximation making use of the Wu-Cohen parametrization~\cite{Wu2006}, that was further checked~\cite{Zhang2023} to provide results totally comparable to PBEsol~\cite{Perdew2008}. The plane wave energy cutoff was of 40 Ha, and the Brillouin zone sampling equivalent to a $8 \times 8 \times 8$ grid for the 5-atom perovskite cubic uc.
The final effective atomic potential contains 360 polynomial terms, until order 8 and was further validated by comparison with first-principles data. It describes well the relative phase stability and distortion amplitudes of the most important metastable phases of \CTO{}. It accurately reproduces the phonon dispersion curves of its $Pnma$ ground state and captures its temperature behavior. It further reproduces the atomic relaxation at ferroelastic twin walls, reproducing results previously obtained from first-principles by Barone et al.~\cite{Barone2014}; see SI Fig. S10. 
Exhaustive details on the construction of the second-principles model of \CTO{} and of its validation by comparison to first-principles calculations, are provided in reference~\cite{Schmitt2020th}.

\section{Acknowledgment}
This work was supported by the Swiss National Science Foundation -- division II -- projects 200020-179155 and 200020-207338, by the Synergia Project N. 154410, and has received funding from the European Research Council under the European Union Seventh Framework Programme (FP7/2007–2013)/ERC Grant Agreement n.319286 (Q-MAC). P.G. acknowledges financial support from F.R.S.-FNRS Belgium (grant PROMOSPAN) and the European Union’s Horizon 2020 research and innovation program under grant agreement number 964931 (TSAR). Calculations were performed on the CECI supercomputer facilities funded by the FRS-FNRS (Grant No. 2.5020.1), the Tier-1 supercomputer of the Fédération Wallonie-Bruxelles funded by the Walloon Region (Grant No. 1117545), and the computing facilities of the Flemish Supercomputer Center. The CIME at EPFL is thanked for access to electron microscope facilities. C. Hébert/LSME of the IPHYS, EPFL are thanked for continued support of D.T.L.A, and for hosting B.M. during much of this work. We thank J. Guo for assistance with plotting PACBED experimental and simulated data, and C. Thibault and C.-Y. Hsu for useful discussions. 

\section{Supporting Information Available}

Supporting Information: Strain states for \LVO{} grown on a $(101)_{\text{orth}}$ \DSO{}; Projections of \textit{Pnma} perovskite structure; STEM HAADF and ABF of \LVO{} film below critical thickness on substrate $[010]_{\text{orth}}$; STEM EELS, EDXS, HAADF and ABF, and PACBED of \LVO{} films above critical thickness; Simulation of PACBED radial distribution functions for bulk \LVO{} and \DSO{}; XRD of $(1\,0\;\frac{1}{2})_\text{pc}$ half-order peak for \LVO{} thickness series; Comparison of first and second principles calculations of \CTO{} twin wall polarization; Structure energy as function of intermediate layer thickness for simulated film thickness series; Image/simulation comparison across switching plane for $[100]_\text{pc}$ substrate zone axis.




\bibliography{main.bib}

\end{document}


\begin{table}[htbp]
\caption{Biaxial epitaxial strain states for \LVO{} grown on a $(101)_{\text{orth}}$ \DSO{} substrate with the film’s orthorhombic long-axis ($b_\text{orth}$) either in-plane ($b_\text{orth} \parallel b_\text{pc}$) or out-of-plane ($b_\text{orth} \parallel c_\text{pc}$), as calculated from bulk lattice constants for the two compounds. $\epsilon_{yy}$ is lower for $b_\text{orth}$ out-of-plane, because the pseudocubic unit cell lattice parameter of \LVO{} is shorter parallel to $b_\text{orth}$ than for the other two pseudocubic axes (themselves defined by $a_\text{orth}$ and $c_\text{orth}$). For this reason, considerations of macroscopic strain energy alone favor film growth with $b_\text{orth}$ out-of-plane.\\}
    \small
    \centering
    \begin{tabular}{ccc}
    \hline
        $b_\text{orth}$ orientation & \hspace{1cm}  & \LVO{} on \DSO{} \\ \hline
        & &$\epsilon_{xx}=0.49$\% \\ 
        $b_\text{orth} \parallel b_\text{pc}$ & & $\epsilon_{yy}=0.71$\%\\ 
        & & $\epsilon_{xy}=0$ \\ \hline       
        & & $\epsilon_{xx}=0.49$\% \\ 
        $b_\text{orth} \parallel c_\text{pc}$ & & $\epsilon_{yy}=0.61$\%\\
        & & $\epsilon_{xy}=0$ \\  \hline
    \end{tabular}
\label{Epitaxial_strain}
\end{table}
\newpage

\begin{figure}[ht]
\begin{center}
\includegraphics[width=\columnwidth]{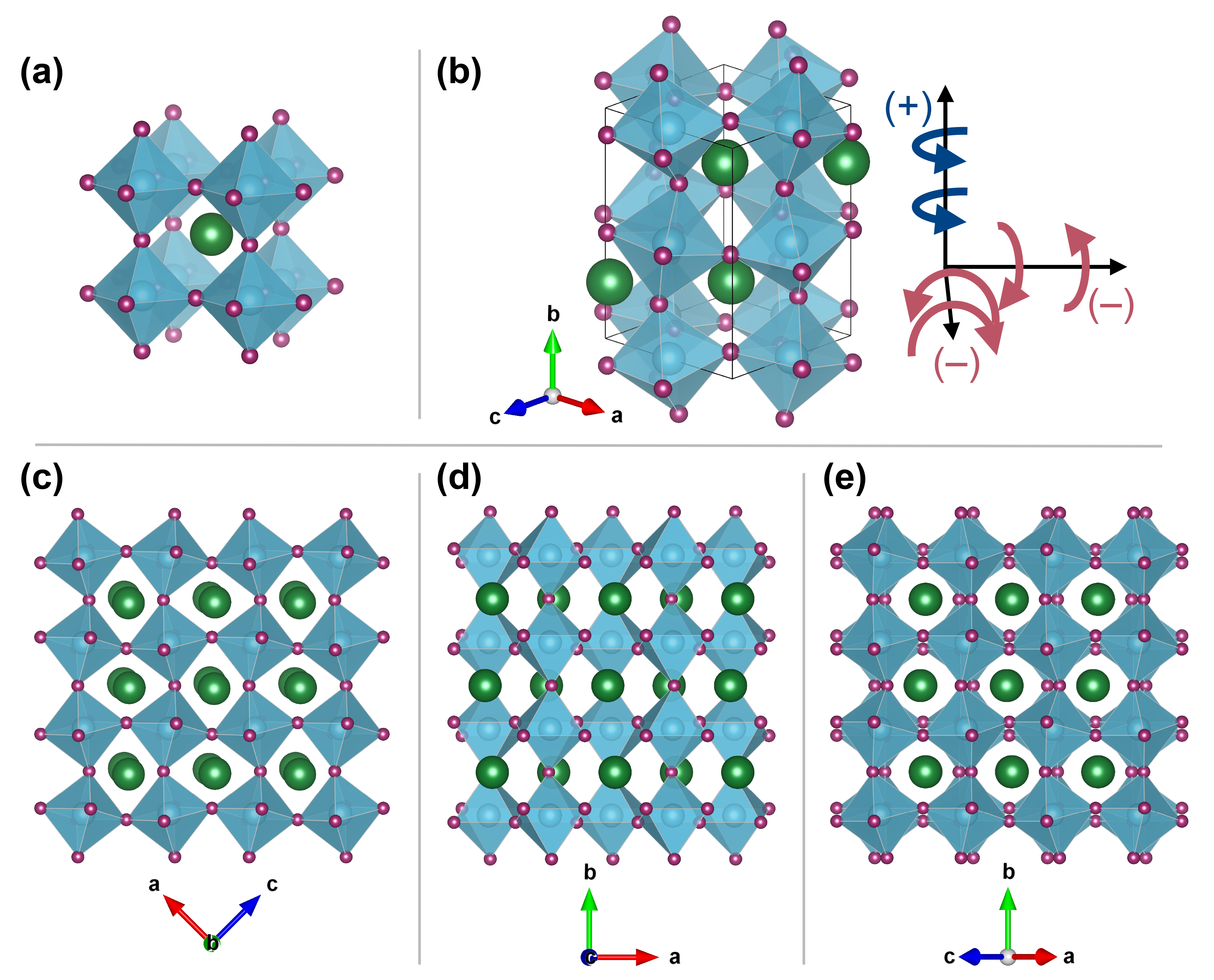}
\caption{The $AB\text{O}_3$ orthorhombic perovskite structure with \textit{Pnma} space group. Panel (a) shows the basic cubic perovskite unit cell, with six oxygen anions (purple spheres) forming an octahedron centered around each corner-sited $B$ cation (blue spheres). The $A$-site cation (green sphere) sits in the middle of four $B\text{O}_6$ octahedra. The unit cell of the orthorhombic perovskite structure is shown in (b). Using the standard \textit{Pnma} setting, this cell has lattice parameters $a_\text{orth} \approx c_\text{orth} \approx \sqrt{2}a_\text{pc}$ and $b_\text{orth} \approx 2a_\text{pc}$  where $a_\text{pc}$ is the lattice parameter of a reference 5-atom pseudocubic unit cell. The $B\text{O}_6$ octahedra of the orthorhombic lattice rotate in-phase along the $b_\text{orth}$  axis and out-of-phase along the two pseudocubic axes that are in the $(010)_\text{orth}$  plane (i.e. plane containing  $a_\text{orth}$  and $c_\text{orth}$); these rotation axes are shown schematically on the right. Projections along different axes highlight the various distortions of this orthorhombic structure: (c) the in-phase oxygen octahedra rotations (OOR) viewed along $[010]_\text{orth}$; (d) the $X_5^-$ mode antipolar motion of the $A$-site cations seen by looking down $[001]_\text{orth}$ (the cations displace parallel to $[100]_\text{orth}$); (e) the out-of-phase OOR seen along $[101]_\text{orth}$/$[100]_\text{pc}$, with this view also showing the $X_5^-$ mode projected by a 45\textdegree~angle. Structural models were prepared with the aid of VESTA~\cite{Momma2011SI}, using an \LVO{} crystal structure file for panels (b)--(e).}
\label{orthorhombic_structure}
\end{center}
\end{figure}

\begin{figure}[ht]
\begin{center}
\includegraphics[width=\columnwidth]{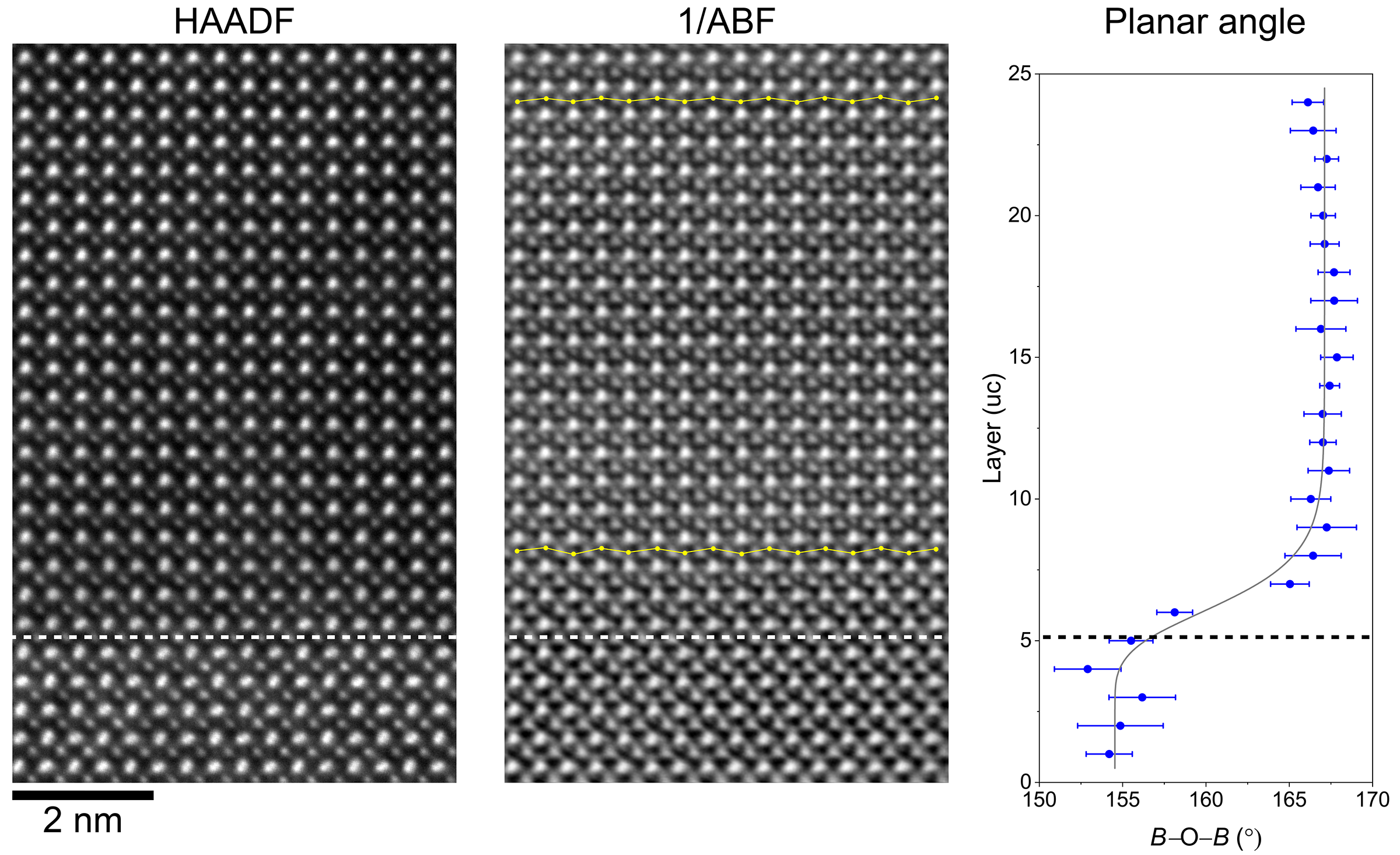}
\caption{Simultaneously-acquired HAADF and (inverted) ABF images of the first layers of a 52 uc \LVO{} film grown on \DSO{}, recorded on the $[010]_\text{orth}$ zone axis of the substrate in order to view the propagation of in-phase OOR from the \DSO{} into the film. The film--substrate interface is indicated with the white dashed line. While data quality is suboptimal, particularly owing to a film–substrate mis-tilt from curvature of the FIB sample upon its preparation, O column positions show sufficient contrast for visualizing the propagation of in-phase OOR from the substrate across the whole film. This is for instance shown in yellow with hand-fitted planar O sites at the 3rd and 19th layers of the film. Atomap has been used to quantify the average planar $B$---O---$B$ angle from layer to layer; the resulting data points on the right confirm the initial reduction in bond angle amplitude going into the film over some uc, followed by continuous propagation at roughly constant amplitude of the in-phase OOR. The results are consistent with the analysis of $A$-site AM presented in Figure 2. The black curve on this plot serves as a guide to the eye, while error bars correspond to the measurement standard deviation across each layer.
}
\label{52uc_ABF}
\end{center}
\end{figure}

\begin{figure}[ht]
\begin{center}
\includegraphics[width=\columnwidth]{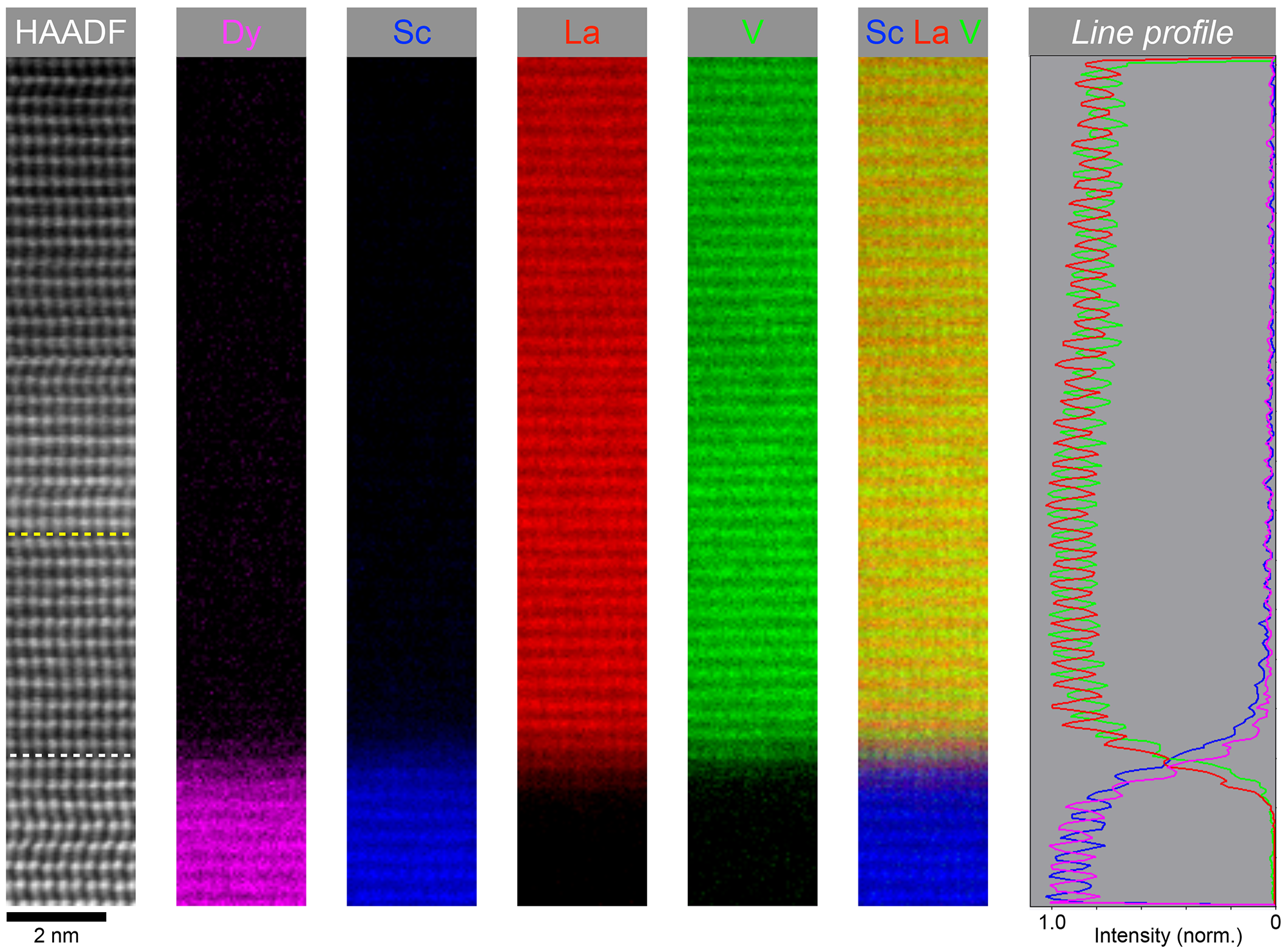}
\caption{STEM-EELS mapping of \LVO{} on \DSO{} at the film--substrate interface of a film grown to thickness $>$ 100 uc, imaged on the $[110]_\text{pc}$ zone axis. HAADF STEM image and associated
Dy $M$, Sc $L$, La $M$ and V $L$ (quantified) maps and RGB composite map are shown. It is seen that the darker contrast in the IL of a HAADF image on this zone axis (indicated between white and yellow dashed lines) is not associated with a change in composition of the IL relative to the film bulk. This result is confirmed by the normalized elemental line profiles on the right, which also show that there are $\sim$2 uc of intermixing at the film--substrate interface. Higher in the film, the elemental line profiles decrease slightly in intensity; this time, the decrease is in accord with the HAADF intensity, indicating that the STEM lamella is reduced in thickness.}
\label{EELS_110pc}
\end{center}
\end{figure}

\begin{figure}[ht]
\begin{center}
\includegraphics[width=\columnwidth]{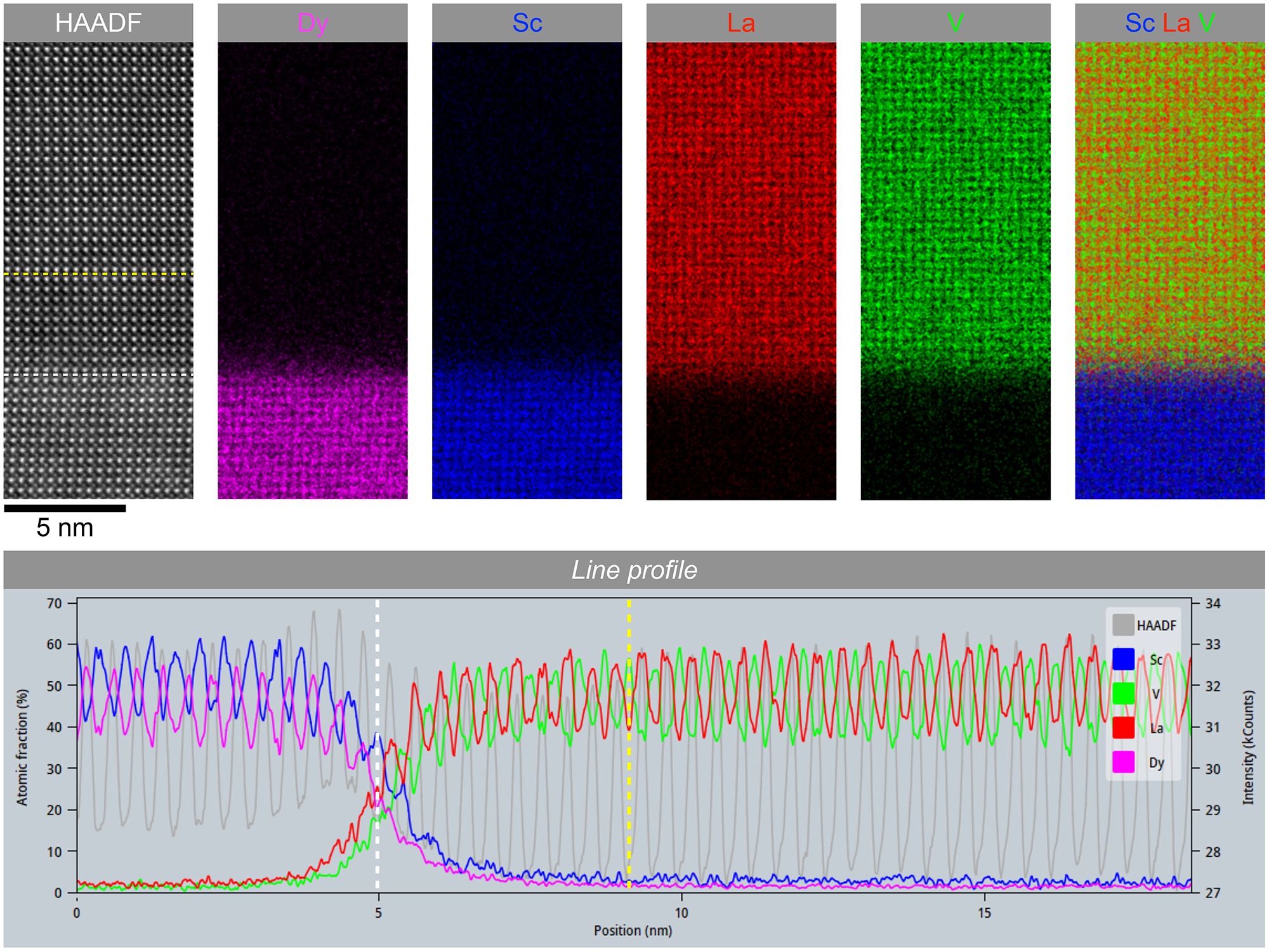}
\caption{HAADF image with associated STEM-EDXS maps and integrated line profile of an 81 uc \LVO{} film grown on \DSO{}, recorded at the film–substrate interface. The maps are recorded along the $[010]_\text{orth}$ zone axis of the substrate (i.e. the orthorhombic long axis). As for SI Fig.~\ref{EELS_110pc}, $\sim$2 uc of chemical intermixing is observed at the film–substrate interface. On this zone axis, the pseudocubic unit cell of the \DSO{} shows a ``shear'' of $\sim$3\textdegree~that is not present in the \LVO{} (whose orthorhombic lattice parameters are nearly tetragonal in nature). The change in angle between the two gives an extra method for identifying the position of the film–substrate interface. This position -- indicated with a white dashed line -- correctly matches the chemically-identified interface, which in turn agrees with the interface position identified from change in HAADF image contrast. (Note that, unlike the $[110]_\text{pc}$ zone axis, the HAADF contrast is not anomalous on this zone axis.) The top of the IL is indicated with a yellow dashed line, as established from the onset of $X_5^-$ antipolar motion of the La cations as, at this position, the \LVO{} zone axis changes from $[010]_\text{orth}$ in the IL to $[10\bar{1}]_\text{orth}$ in the film bulk. Atomic fractions in the line profile are calculated only with respect to the cations.}
\label{EDXS_001pc}
\end{center}
\end{figure}

\begin{figure}[ht]
\begin{center}
\includegraphics[width=\columnwidth]{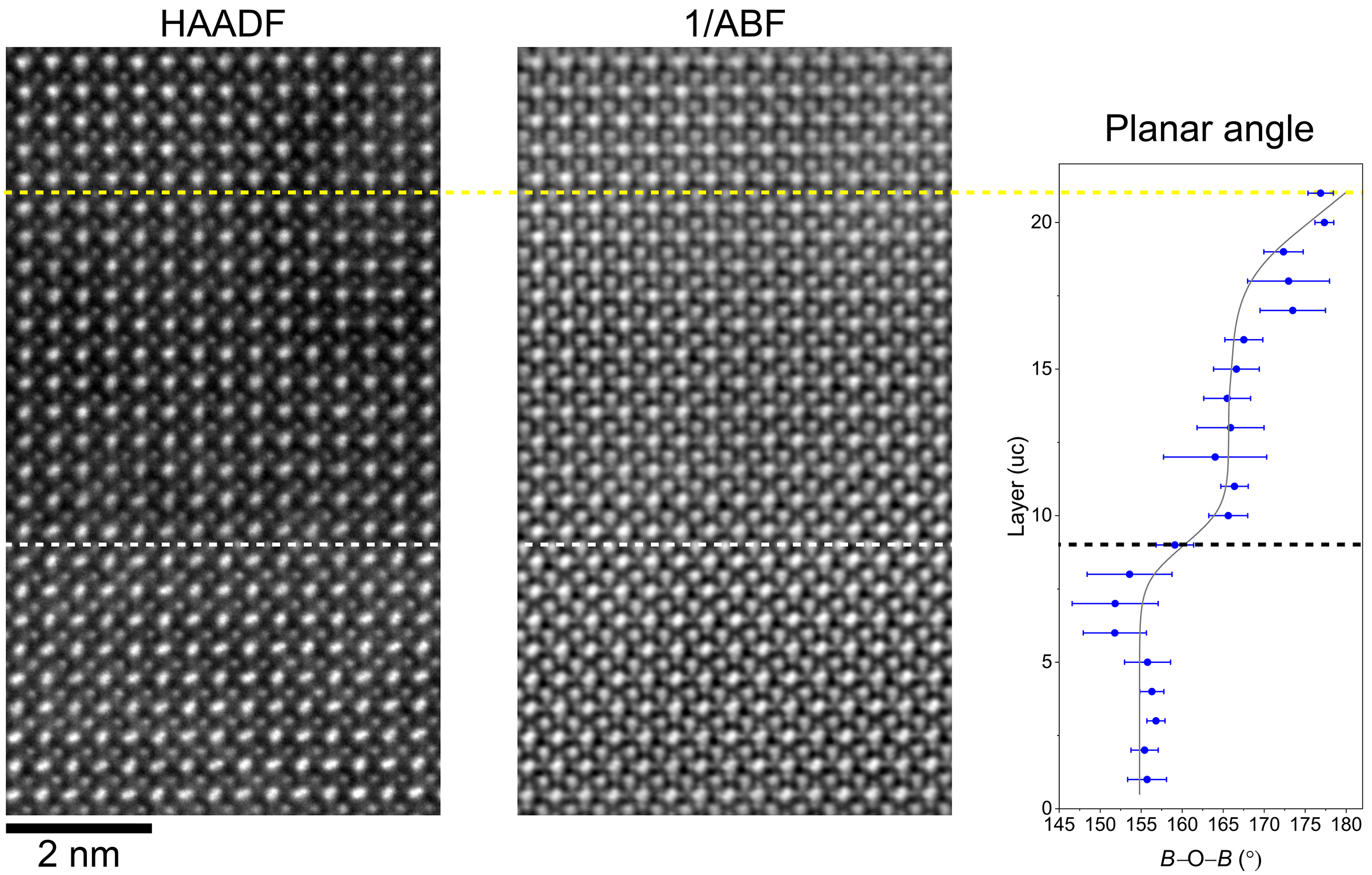}
\caption{Simultaneously-acquired HAADF and (inverted) ABF images of IL of 81 uc \LVO{} grown on \DSO{} recorded on the $[010]_\text{orth}$ zone axis of the substrate, in order to view the propagation of in-phase OOR from the \DSO{} into the IL. The film–substrate interface is indicated with the white dashed line, and the switching plane position with the yellow dashed line (as evaluated from the onset of $X_5^-$ mode at the top). While data quality is suboptimal owing to experimental challenges -- including a film--substrate mis-tilt from curvature of the FIB sample upon its preparation -- the O column positions show sufficient contrast for visualizing the trend of $B$---O---$B$ angles. Atomap has been used to quantify the average planar $B$---O---$B$ angle from layer to layer. Results show the same trend as for the AM measurement in Fig. 4: initial reduction of orthorhombic distortion (corresponding to an increase in bond angle amplitude) at the film–substrate interface indicated by the black dashed lines, followed by a plateau region at $\sim$165\textdegree, and then tending to a value of $\sim$180\textdegree~at the top of the IL. The curve in black acts as a guide to the eye. Note that positions are not measured above the switching plane, because the planar O site becomes a projected dumbbell as the in-phase rotation axis switches to out-of-plane. In this region of the deposited film, the IL appears 1--2 uc thicker than the region shown in Fig. 4. This is suggested to derive from a small instability in the formation energetics, that is discussed in the main text.}
\label{81uc_ABF}
\end{center}
\end{figure}

\begin{figure}[ht]
\begin{center}
\includegraphics[width=\columnwidth]{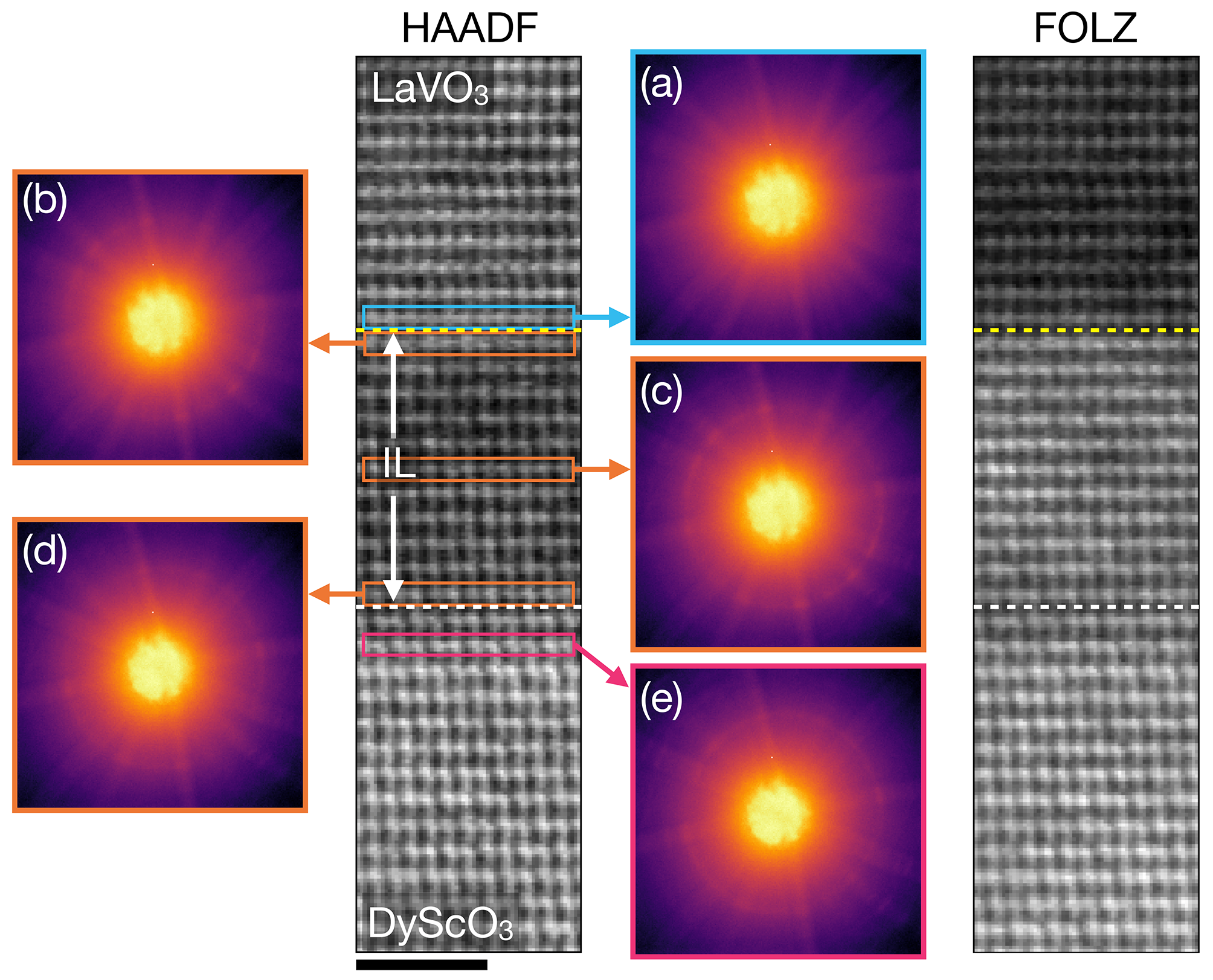}
\caption{PACBED measurements of an 81 uc film of \LVO{} grown on \DSO{}, recorded along the $[110]_\text{pc}$/$[11\bar{1}]_\text{orth}$  zone axis of the substrate. Each displayed pattern corresponds to the average STEM diffraction pattern from the associated 1 uc thick region shown on the HAADF STEM image. Patterns (b), (c) and (d) from the \LVO{} IL maintain the characteristic low-angle first order Laue zone (FOLZ) of the \DSO{} substrate that is seen in pattern (e). Stepping just one unit cell from (b) to (a) across the switching plane (indicated with a yellow dashed line) leads to a sharp loss of this FOLZ ring as the film discretely switches orientation from a $[11\bar{1}]_\text{orth}$  to a $[001]_\text{orth}$  zone axis. The data were acquired using a ``4D-STEM'' approach, recording a diffraction pattern at each probe position, with the microscope operated at a high tension of 200 kV (unlike 300 kV for the other measurements). The displayed HAADF image was generated by integrating the 4D-STEM signal over scattering angles from 80--140 mrad. In order to demonstrate the significance of the low-angle FOLZ, a FOLZ image on the right is generated by integrating the signal intensity across the FOLZ ring over scattering angles of 64--70 mrad. Above the switching plane, the FOLZ image intensity drops sharply because of the disappearance of the FOLZ ring. Scale bar: 2 nm.}
\label{PACBED_4DSTEM}
\end{center}
\end{figure}

\begin{figure}[ht]
\begin{center}
\includegraphics[width=\columnwidth]{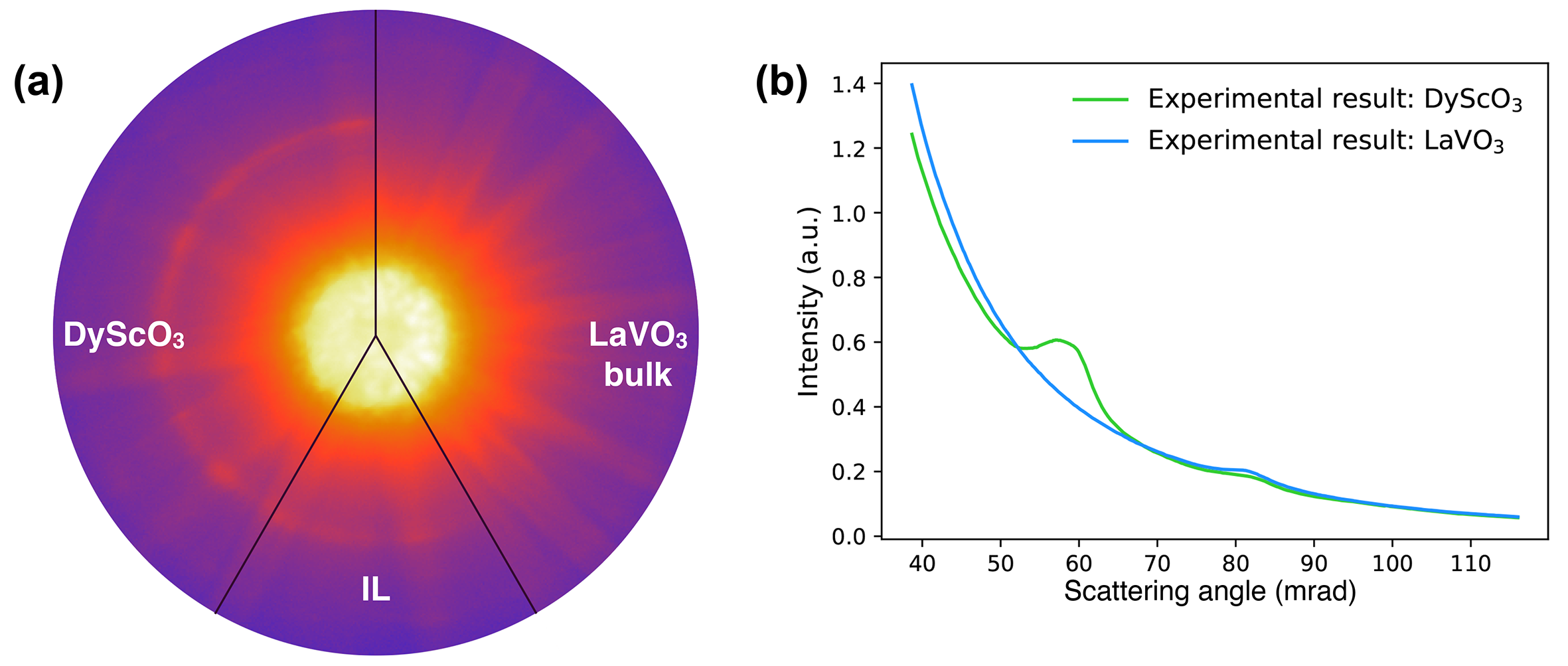}
\caption{Anomalous HAADF intensity along the $[11\bar{1}]_\text{orth}$ zone axis: experimental data.
(a) PACBED patterns from Fig. 3, arranged into a composite pattern for easy comparison.
(b) From the \DSO{} substrate and \LVO{} film bulk PACBED patterns, experimental radial distribution functions (RDFs) of the signal intensity versus scattering angle are determined. The substrate, having a $[11\bar{1}]_\text{orth}$ zone axis shows a strong peak at $\sim$60 mrad scattering angle, for a low angle FOLZ ring. This peak is not present in the film bulk, which instead has a $[001]_\text{orth}$ zone axis. Outside of this peak, the \LVO{} actually has a higher intensity than the substrate, even though it has a lower average atomic number ($Z_\text{av}$) of 19.6 \textit{vs.} 21, that typically leads to decreased intensity from thermal diffuse scattering (intensity $\propto Z_\text{av}^{1.6-2}$). It is suggested that, on the substrate’s $[11\bar{1}]_\text{orth}$  zone axis, strong coherent elastic scattering into the low-angle FOLZ leads to an anomalously low thermal diffuse scattering (TDS), from a partitioning effect on the total scattering.}
\label{RDF_exp}
\end{center}
\end{figure}

\begin{figure}[ht]
\begin{center}
\includegraphics[width=\columnwidth]{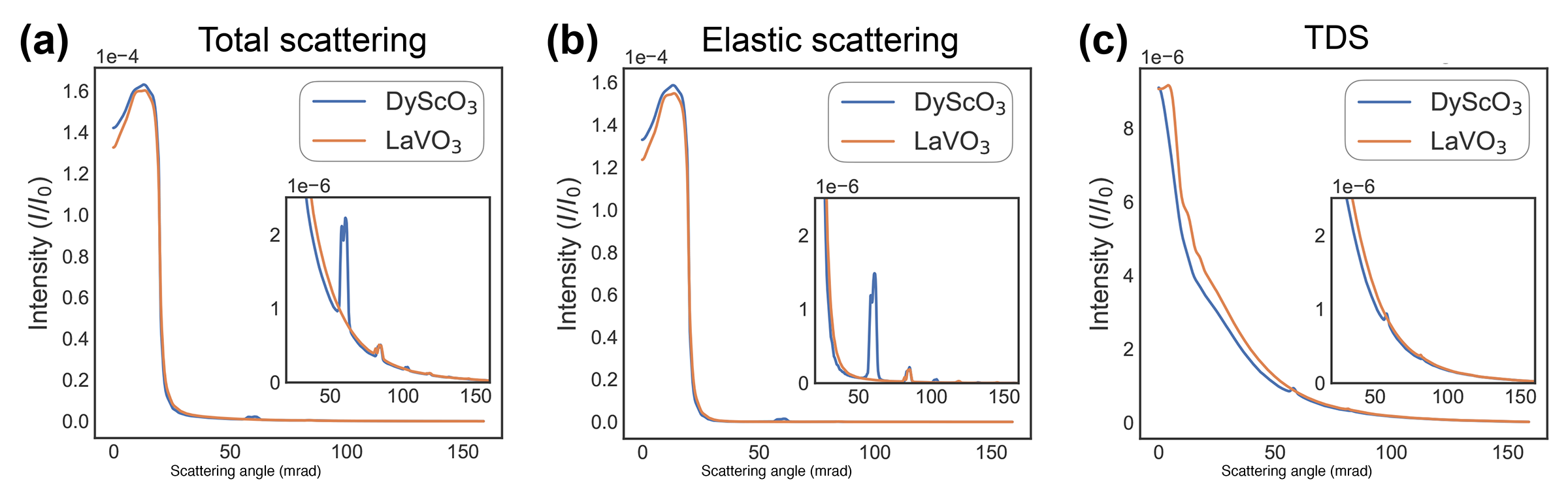}
\caption{Anomalous HAADF intensity along the $[11\bar{1}]_\text{orth}$ zone axis: RDF simulations.
The quantum excitation of phonons model of the $\mu$STEM simulation software is applied, allowing the incoherent inelastic scattering of phonon-driven TDS to be separated from coherent elastic scattering~\cite{Forbes2010SI,Allen2015SI,MuSTEM}. $\mu$STEM was used to simulate PACBED patterns, from which RDFs were calculated. (a), (b) and (c) respectively show the total scattering, elastic scattering and TDS RDFs for substrate and bulk film. The total scattering simulations confirm that, outside of the FOLZ, the \DSO{} has a lower scattering intensity than the \LVO{}, as seen experimentally in SI Fig.~\ref{RDF_exp}. Moreover, the TDS curves confirm that this is specifically because the \DSO{} has an anomalously low thermal diffuse scattering -- lower than that for the \LVO{}, despite the higher $Z_\text{av}$ of \DSO{}. Therefore, increased coherent elastic scattering into the FOLZ leads to a net reduction of phonon-driven TDS via a partitioning effect. This result in turn explains the darker HAADF contrast of the \LVO{} IL compared to the film bulk, when the sample is imaged on the $[110]_\text{pc}$ zone axis (Fig. 3). The IL keeps the $[11\bar{1}]_\text{orth}$ zone axis of the substrate. It therefore also suffers from an anomalously low thermal diffuse scattering. By using a high HAADF collection angle that excludes the FOLZ, the HAADF signal from the IL is then lower in intensity than the bulk \LVO{}, giving rise to its darker contrast in the HAADF STEM image.}
\label{RDF_sim}
\end{center}
\end{figure}

\begin{figure}[ht]
\begin{center}
\includegraphics[width=0.6\columnwidth]{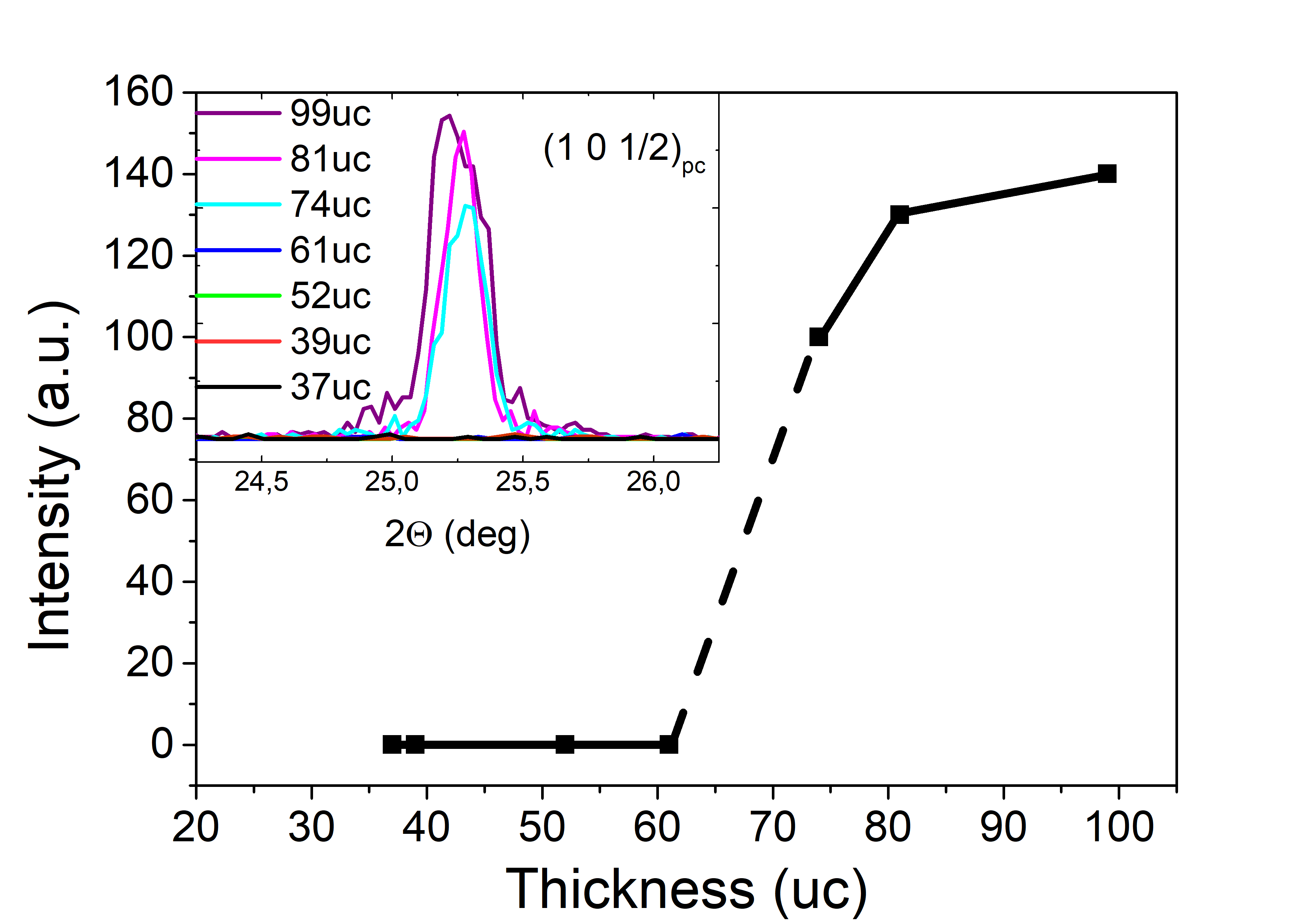}
\caption{The measurement by XRD of the $(1\,0\;\frac{1}{2})_\text{pc}$ half-order peak (see inset) reveals the presence of the in-phase $(+)$ OOR axis ($b_\text{orth}$) in the out-of-plane direction: the intensity of this reflection becomes non-zero only for \LVO{} films with a thickness above a critical value of 62--74~uc.}
\label{XRD}
\end{center}
\end{figure}

\begin{figure}[ht]
\begin{center}
\includegraphics[width=0.6\columnwidth]{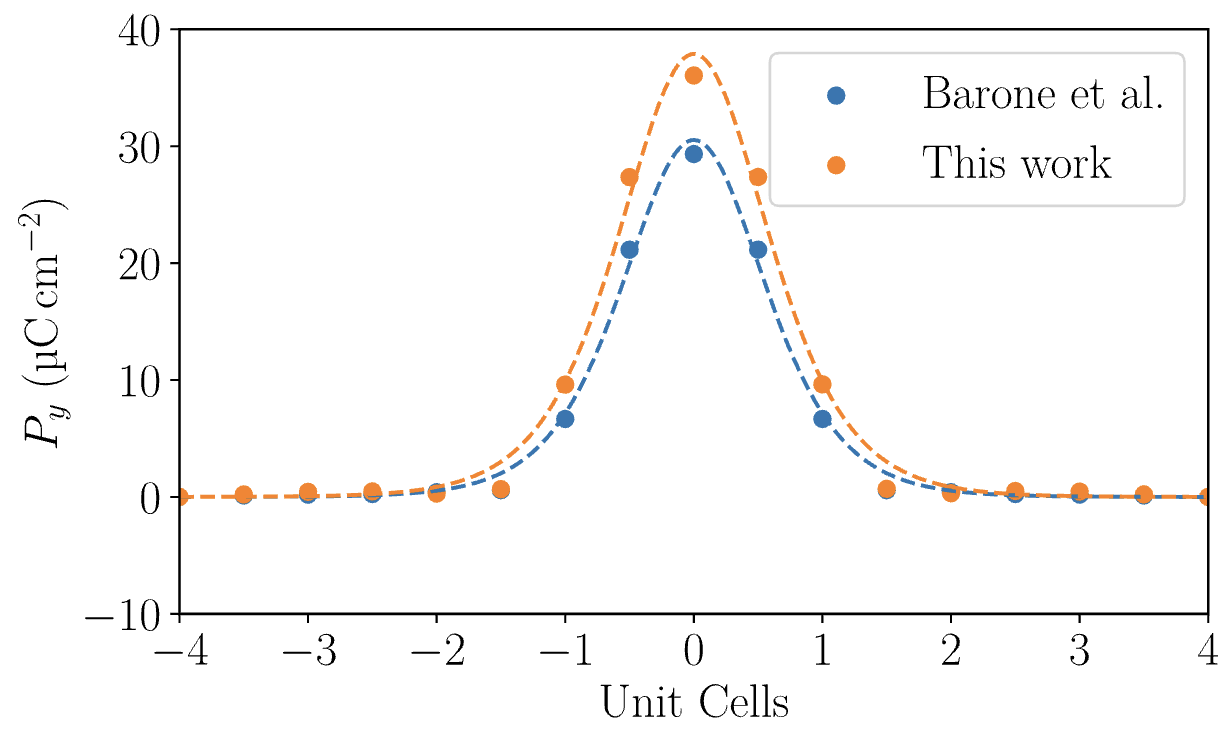}
\caption{Comparison of the evolution of the polarization at a twin wall of \CTO{} as obtained from our second-principles model and from first-principles calculations by Barone et al.; see Fig. 2 of Ref.~\cite{Barone2014SI}.
}
\label{twin_wall}
\end{center}
\end{figure}

\begin{figure}[ht]
\begin{center}
\includegraphics[width=0.8\columnwidth]{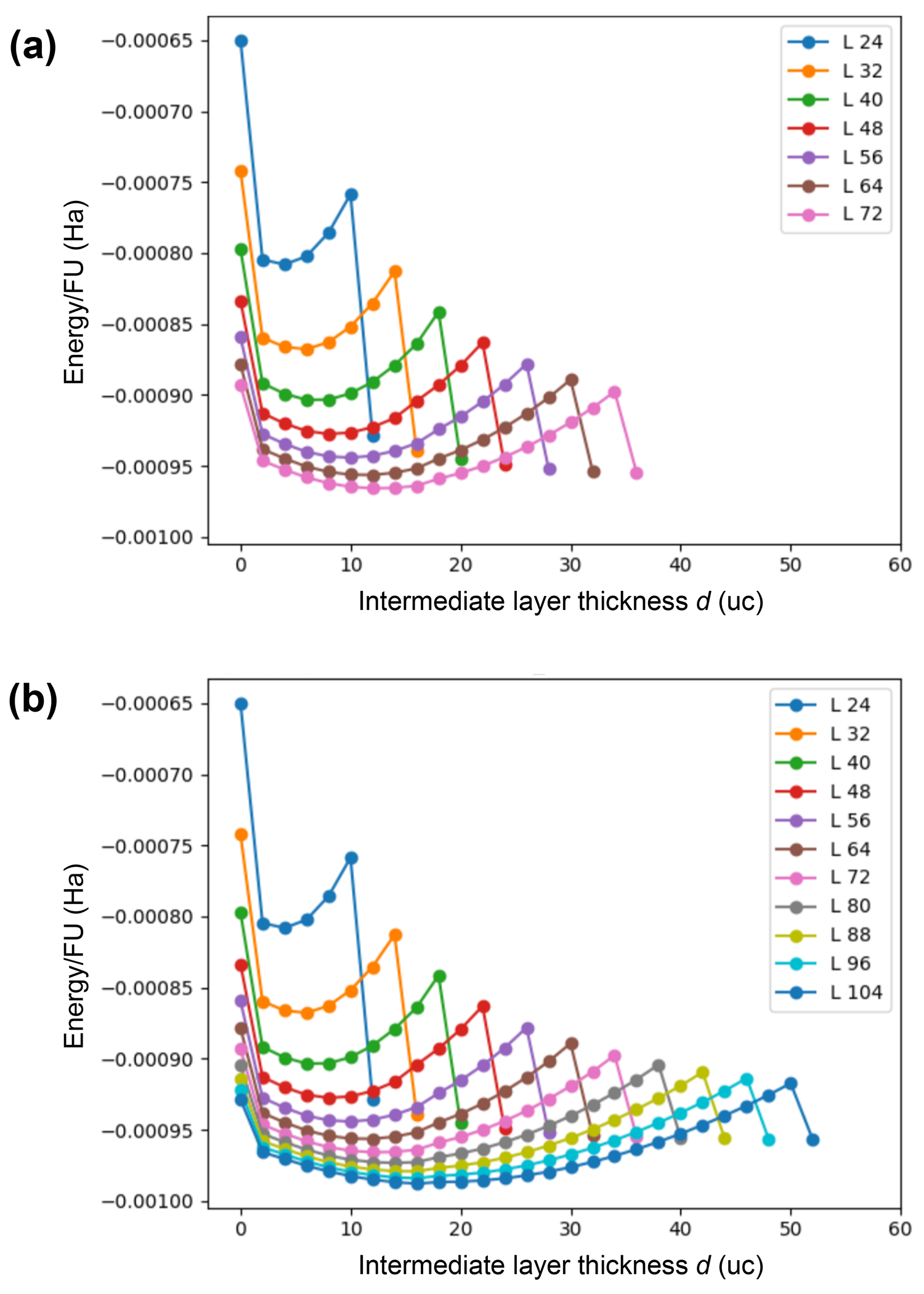}
\caption{From the second-principles calculations, structure energy as a function of IL thickness $d$ for supercells ranging in uc dimension from (a) $L=24$ to $L=72$ and (b) $L=24$ to $L=104$.}
\label{Energetic_curves}
\end{center}
\end{figure}

\begin{figure}[ht]
\begin{center}
\includegraphics[width=\columnwidth]{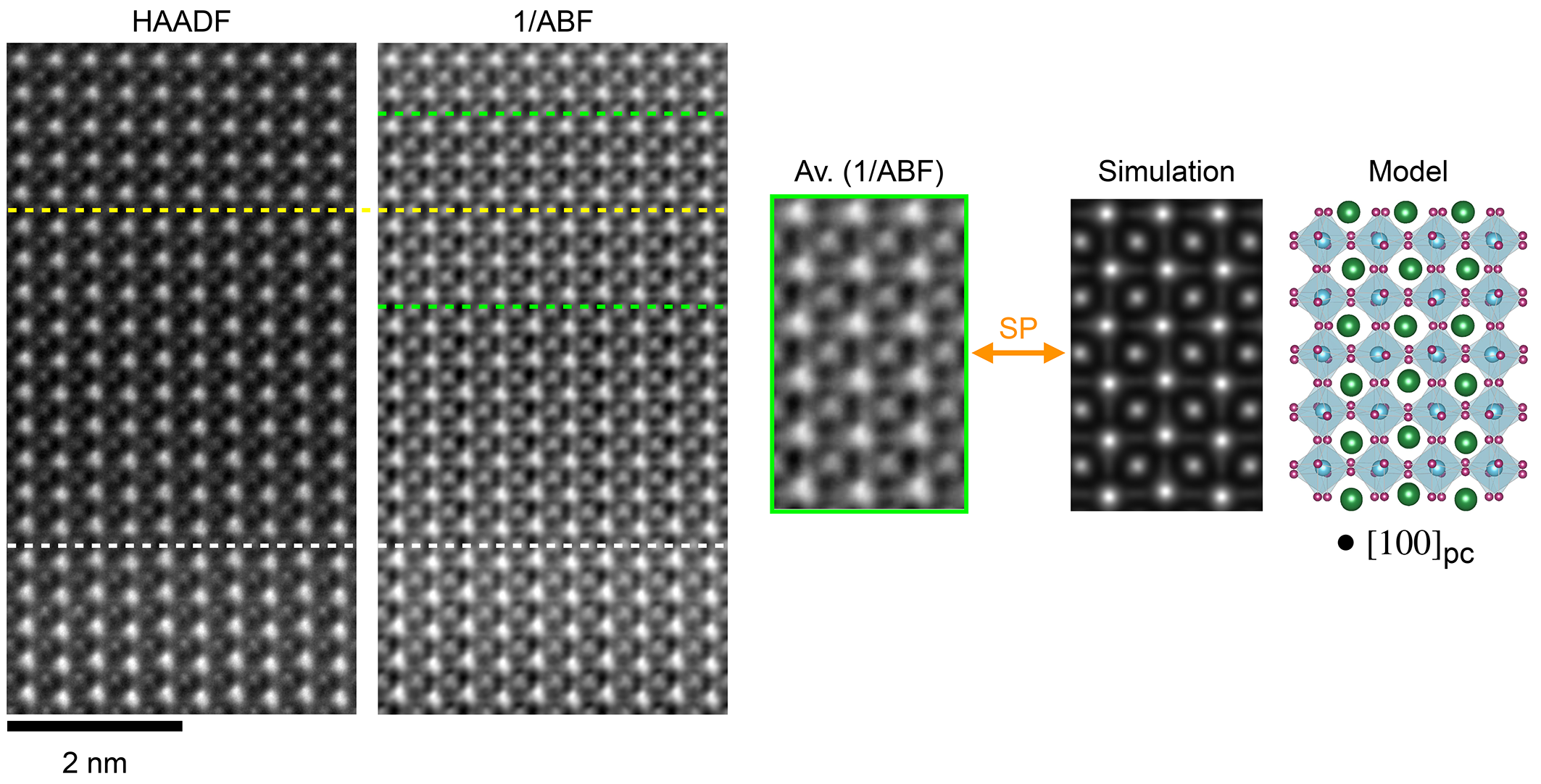}
\caption{On the left, simultaneously-acquired HAADF and (inverted) ABF images of IL of 81 uc \LVO{} grown on \DSO{}, recorded on the $[100]_\text{pc}$/$[10\bar{1}]_\text{orth}$ zone axis of the substrate. The ABF image has been treated with a Wiener filter to reduce noise. The white and yellow dashed lines respectively indicate the film--substrate interface and the switching plane. From the full 20 uc width of the initial ABF image, subframe averaging using the SmartAlign template-matching module has been applied. From this averaged image, the segment around the switching plane is shown in the middle ``Av. (1/ABF)'' image, whose vertical positioning corresponds to the green dashed lines on the non-averaged 1/ABF image. This experimental image is compared to a simulated image based on the second-principles modeling-determined structure on the right. The streaks between the $A$-sites derive from the O dumbbells that themselves are indicative of the out-of-phase OOR on this $[100]_\text{pc}$ zone axis. In the model, it is seen that this OOR mode propagates across the switching plane (SP) with negligible damping. By comparing the experimental image to the simulated one, it is seen that the experimental image is consistent with this finding.}
\label{81uc_100pc-ABF}
\end{center}
\end{figure}

\FloatBarrier
\bibliography{main.bib}